\PassOptionsToPackage{prologue,usenames,dvipsnames,table}{xcolor}
\PassOptionsToPackage{bookmarks,unicode,colorlinks=true}{hyperref}

\newif\ifanonymous

\anonymousfalse
\ifanonymous
\documentclass[acmsmall,screen,review,anonymous]{acmart}
\else
\documentclass[acmsmall,screen]{acmart}
\fi

\usepackage[ruled,vlined,linesnumbered]{algorithm2e}
\SetArgSty{textrm}
\SetKwComment{tcp}{$\rhd\;$}{}%
\makeatletter
\patchcmd{\@algocf@start}% <cmd>
{-1.5em}% <search>
{0pt}% <replace>
{}{}% <success><failure>
\makeatother

\makeatother
\usepackage{threeparttable}
\usepackage{todonotes,xspace}
\usepackage{xcolor}
\definecolor{shadecolor}{rgb}{0.90, 0.90, 0.90}

\newcommand{\green}[1]{\textcolor{ForestGreen}{#1}}

\newcommand{\red}[1]{\textcolor{Maroon}{#1}}

\usepackage[many]{tcolorbox}

\definecolor{promptbg}{RGB}{245,245,245}    % 浅灰色背景
\definecolor{titlebg}{RGB}{235,235,235}      % 稍深的标题背景

\newtcolorbox{promptbox}[1]{
    enhanced,
    breakable,
    colback=promptbg,        % 内容区背景色
    colbacktitle=titlebg,    % 标题区背景色
    colframe=titlebg,
    coltitle=black,          % 标题文字颜色
    fonttitle=\bfseries\text\small,
    fontupper=\text\small,
    title=#1,               % 标题参数
    boxrule=1.5pt,          % 边框宽度
    arc=1pt,                % 圆角半径
    left=10pt,              % 左内边距
    right=10pt,             % 右内边距
    top=4pt,                % 上内边距
    bottom=2pt,             % 下内边距
    toptitle=1pt,           % 标题上间距
    bottomtitle=1pt,        % 标题下间距
    titlerule=0.5pt,        % 标题分隔线宽度
    titlerule style={titlebg} % 标题分隔线颜色
}

\usepackage{graphicx}
\usepackage{caption}
\usepackage{subcaption}
\usepackage{amsmath}
\usepackage{thm-restate}

\usepackage{circledsteps}

\renewcommand{\Circled}[1]{\textcircled{\raisebox{-.87pt}{\texttt{#1}}}}

\usepackage{mathtools}
\usepackage{csquotes}

\usepackage{fancyvrb}
\usepackage[frozencache,cachedir=minted-cache]{minted} % for ArXiv submission
\setminted[c]{
    frame=lines,
    framesep=2mm,
    baselinestretch=1.0,
    fontsize=\footnotesize, % a bit larger
    breaklines=true,
    xleftmargin=0em,
    xrightmargin=0em,
    autogobble,
    escapeinside=||, % 用于特殊注释处理
    showspaces=false,
    showtabs=false
}

\usepackage{framed}

\usepackage{adjustbox}
\usepackage{makecell}
\usepackage{tikz}
\usetikzlibrary{patterns,calc,arrows,shapes,snakes,automata,backgrounds,petri,positioning,decorations.pathmorphing,intersections}
\usepackage{tkz-euclide}
\tikzset{mynode/.style={draw,solid,circle,inner sep=1pt}}
\usepackage[customcolors]{hf-tikz}
\hfsetfillcolor{gray!10}
\hfsetbordercolor{none}
\usepackage{pgfplots}
\usepgfplotslibrary{units}  
\pgfplotsset{compat=1.17}
\usepackage{multirow}
\usepackage{pifont} 

\usepackage{hyperref}
\usepackage[
    type={CC},
    modifier={by-nc-sa},
    version={3.0},
]{doclicense}

\usepackage[capitalise]{cleveref}
\crefname{equation}{}{}
\crefname{figure}{Fig.}{Figs.}
\crefdefaultlabelformat{#2\textup{#1}#3}

\newcommand{\vars}{\mathtt{vars}}
\newcommand{\vals}{\mathtt{vals}}
\newcommand{\pstate}{\ensuremath{\sigma}\xspace}
\newcommand{\States}{\mathit{\Sigma}}

\newcommand{\cgraph}{\ensuremath{\mathcal{G}}\xspace}

\newcommand{\ours}{\textup{\textsc{Preguss}}\xspace}

\newcommand{\eva}{\textup{\textsc{Frama-C/Eva}}\xspace}
\newcommand{\rte}{\textup{\textsc{Frama-C/Rte}}\xspace}
\newcommand{\Wp}{\textup{\textsc{Frama-C/Wp}}\xspace}

\newcommand{\autospec}{\textup{\textsc{AutoSpec}}\xspace}
\newcommand{\autospecRte}{$\textup{\textsc{AutoSpec}}_\textup{\textsc{Rte}}$\xspace}

 \newcommand{\true}{\ensuremath{\textnormal{\texttt{true}}}}

 \newcommand{\tagTrue}{\ensuremath{\textit{\textcolor{ForestGreen}{true}}}}
 \newcommand{\tagUnknown}{\ensuremath{\textit{\textcolor{Maroon}{unknown}}}}

\newcommand{\aand}{\ \ \textnormal{and}\ \ }

\newcommand{\qimplies}{\ \quad\textnormal{implies}\quad\ }

\newcommand{\defeq}{{}\triangleq{}}
\newcommand{\ddefeq}{\ {}\defeq{}\ }
\newcommand{\eeq}{~{}={}~}

\newcommand{\mmmapsto}{~{}\mapsto{}~}

\newcommand{\mmodels}{~{}\models{}~}

\def\triangleforqed{\hbox{$\lhd$}}
\makeatletter
\DeclareRobustCommand{\qedT}{%
	\ifmmode
	\eqno \def\@badmath{$$}%$$
	\let\eqno\relax \let\leqno\relax \let\veqno\relax
	\hbox{\triangleforqed}%
	\else
	\leavevmode\unskip\penalty9999 \hbox{}\nobreak\hfill
	\quad\hbox{\triangleforqed}%
	\fi
}
\makeatother

\theoremstyle{plain}
\newtheorem*{thm*}{Theorem}
\newtheorem*{exmp*}{Example}
\newtheorem*{counterexmp*}{Counterexample}

\newtheoremstyle{ourstyle}{}{}{\itshape}{}{\bfseries}{.}{ }{\thmname{#1}\thmnumber{ #2}\thmnote{ (#3)}}
\theoremstyle{ourstyle}

\newtheorem{theorem}{Theorem}
\newtheorem{definition}[theorem]{Definition}
\newtheorem{lemma}[theorem]{Lemma}

\newtheoremstyle{exmpstyle}{}{}{}{}{\bfseries}{.}{ }{\thmname{#1}\thmnumber{ #2}\thmnote{ (#3)}}
\theoremstyle{exmpstyle}

\newtheorem{example}[theorem]{Example}

\newtheoremstyle{rmkstyle}{}{}{}{}{\bfseries}{.}{ }{\thmname{#1}\thmnote{ (#3)}}
\theoremstyle{rmkstyle}

\newtheorem{remark}{Remark}

\newcounter{challenge}
\renewcommand{\thechallenge}{\arabic{challenge}} % change formatting if you want section-based numbering, e.g. \thesection.\arabic{challenge}

\crefname{challenge}{Challenge}{Challenges}   % lower-case for \cref
\Crefname{challenge}{Challenge}{Challenges}   % capitalized for \Cref

\newenvironment{challenge}[1][]% optional argument passed to tcolorbox
{%
  \refstepcounter{challenge}% makes the counter targettable by \label and hyperref
  \begin{tcolorbox}[boxrule=1pt,colback=white,colframe=black!75,boxsep=0mm,#1]%
  \textbf{Challenge \thechallenge.}\ %
}
{%
  \end{tcolorbox}%
}

\newcounter{insight}
\renewcommand{\theinsight}{\arabic{insight}} % change formatting if you want section-based numbering, e.g. \thesection.\arabic{insight}

\crefname{insight}{Insight}{Insights}   % lower-case for \cref
\Crefname{insight}{Insight}{Insights}   % capitalized for \Cref

\newenvironment{insight}[1][]% optional argument passed to tcolorbox
{%
  \refstepcounter{insight}% makes the counter targettable by \label and hyperref
  \begin{tcolorbox}[boxrule=1pt,colback=white,colframe=black!75,boxsep=0mm,#1]%
  \textbf{Insight \theinsight.}\ %
}
{%
  \end{tcolorbox}%
}

\providecommand*\phantomsection{}

\AtBeginDocument{%
  }

\makeatletter
\renewcommand\paragraph{\@startsection{paragraph}{4}{\z@}%
  {-.5\baselineskip \@plus -2\p@ \@minus -.2\p@}%
  {-3.5\p@}%
  {\ACM@NRadjust{\@parfont\@adddotafter}}}
\makeatother

\setcopyright{none}
\acmDOI{10.1145/3798268}
\acmYear{2026}
\acmJournal{PACMPL}
\acmVolume{10}
\acmNumber{OOPSLA1}
\acmArticle{160}
\acmMonth{4}
\received{2025-10-10}
\received[accepted]{2026-02-17}

\begin{document}

% \title{{\preguss}: Potential Runtime Error-Guided Specification Synthesis}
\title{A Tale of 1001 LoC: Potential Runtime Error-Guided Specification Synthesis for Verifying Large-Scale Programs}

\author{Zhongyi Wang}
\orcid{0009-0008-1986-6070}
\authornote{Both authors contributed equally to this research.}
\affiliation{%
  \institution{Zhejiang University}
  \city{Hangzhou}
  \country{China}}
\email{zhongyi.wang@zju.edu.cn}

\author{Tengjie Lin}
\orcid{0009-0002-5951-7314}
\authornotemark[1]
\affiliation{%
  \institution{Zhejiang University}
  \city{Hangzhou}
  \country{China}}
\email{tengjie.lin@zju.edu.cn}

\author{Mingshuai Chen}
\authornote{Corresponding author.}
\orcid{0000-0001-9663-7441}
\affiliation{%
  \institution{Zhejiang University}
  \city{Hangzhou}
  \country{China}}
\email{m.chen@zju.edu.cn}

\author{Haokun Li}
\orcid{0000-0001-6411-9324}
\affiliation{%
  \institution{Peking University}
  \city{Beijing}
  \country{China}}
\email{ker@pm.me}

\author{Mingqi Yang}
\orcid{0009-0004-5304-6763}
\affiliation{%
  \institution{Zhejiang University}
  \city{Hangzhou}
  \country{China}}
\email{mingqiyang@zju.edu.cn}

\author{Xiao Yi}
\orcid{0000-0002-4792-4433}
\affiliation{%
  \institution{The Chinese University of Hong Kong}
  \city{Hong Kong}
  \country{China}}
\email{yixiao5428@link.cuhk.edu.hk}

\author{Shengchao Qin}
\orcid{0000-0003-3028-8191}
\affiliation{%
  \institution{Xidian University}
  \city{Xi'an}
  \country{China}}
\email{shengchao.qin@gmail.com}

\author{Yixing Luo}
\orcid{0000-0001-6223-5217}
\affiliation{%
  \institution{Beijing Institute of Control Engineering}
  \city{Beijing}
  \country{China}}
\email{luoyi_xing@126.com}

\author{Xiaofeng Li}
\orcid{0009-0002-8805-665X}
\affiliation{%
  \institution{Beijing Institute of Control Engineering}
  \city{Beijing}
  \country{China}}
\email{li_x_feng@126.com}

\author{Bin Gu}
\orcid{0000-0001-7218-4839}
\affiliation{%
  \institution{Beijing Institute of Control Engineering}
  \city{Beijing}
  \country{China}}
\email{gubin@ios.ac.cn}

\author{Liqiang Lu}
\orcid{0000-0002-3801-6847}
\affiliation{%
  \institution{Zhejiang University}
  \city{Hangzhou}
  \country{China}}
\email{liqianglu@zju.edu.cn}

\author{Jianwei Yin}
% \authornotemark[2]
\orcid{0000-0003-4703-7348}
\affiliation{%
  \institution{Zhejiang University}
  \city{Hangzhou}
  \country{China}}
\email{zjuyjw@zju.edu.cn}

\renewcommand{\shortauthors}{Z Wang, T Lin, M Chen, H Li, M Yang, X Yi, S Qin, Y Luo, X Li, B Gu, L Lu and J Yin}

\begin{abstract}
% Fully automated verification of large-scale software and hardware systems is arguably the holy grail of formal methods. Large language models (LLMs) have recently demonstrated their potential for enhancing the degree of automation in formal verification by, e.g., generating formal specifications as essential to deductive verification, yet exhibit poor scalability due to context-length limitations and, more importantly, the difficulty of inferring complex, interprocedural specifications. This paper presents \textsc{Preguss} -- a modular, fine-grained framework for automating the generation and refinement of formal specifications. \textsc{Preguss} synergizes between static analysis and deductive verification by steering two components in a divide-and-conquer fashion: (i) potential runtime error-guided construction and prioritization of verification units, and (ii) LLM-aided synthesis of interprocedural specifications at the unit level. We show that {\ours} substantially outperforms state-of-the-art LLM-based approaches and, in particular, it enables highly automated RTE-freeness verification for real-world programs with over a thousand LoC, with a reduction of 80.6\%$\sim$88.9\% human verification effort.

Fully automated verification of large-scale software and hardware systems is arguably the holy grail of formal methods. Large language models (LLMs) have recently demonstrated their potential for enhancing the degree of automation in formal verification by, e.g., generating formal specifications as essential to deductive verification, yet exhibit poor scalability due to long-context reasoning limitations and, more importantly, the difficulty of inferring complex, interprocedural specifications. This paper presents \textsc{Preguss} -- a modular, fine-grained framework for automating the generation and refinement of formal specifications. \textsc{Preguss} synergizes between static analysis and deductive verification by steering two components in a divide-and-conquer fashion: (i) potential runtime error-guided construction and prioritization of verification units, and (ii) LLM-aided synthesis of interprocedural specifications at the unit level. We show that {\ours} substantially outperforms state-of-the-art LLM-based approaches and, in particular, it enables highly automated RTE-freeness verification for real-world programs with over a thousand LoC, with a reduction of 80.6\%$\sim$88.9\% human verification effort.
\end{abstract}


\begin{CCSXML}
<ccs2012>
   <concept>
       <concept_id>10003752.10010124.10010138.10010142</concept_id>
       <concept_desc>Theory of computation~Program verification</concept_desc>
       <concept_significance>500</concept_significance>
       </concept>
   <concept>
       <concept_id>10003752.10010124.10010138.10010140</concept_id>
       <concept_desc>Theory of computation~Program specifications</concept_desc>
       <concept_significance>500</concept_significance>
       </concept>
   <concept>
       <concept_id>10003752.10010124.10010138.10010143</concept_id>
       <concept_desc>Theory of computation~Program analysis</concept_desc>
       <concept_significance>300</concept_significance>
       </concept>
   <concept>
       <concept_id>10011007.10011074.10011099.10011692</concept_id>
       <concept_desc>Software and its engineering~Formal software verification</concept_desc>
       <concept_significance>500</concept_significance>
       </concept>
   <concept>
       <concept_id>10011007.10010940.10010992.10010998.10011000</concept_id>
       <concept_desc>Software and its engineering~Automated static analysis</concept_desc>
       <concept_significance>300</concept_significance>
       </concept>
 </ccs2012>
\end{CCSXML}

\ccsdesc[500]{Theory of computation~Program verification}
\ccsdesc[500]{Theory of computation~Program specifications}
\ccsdesc[300]{Theory of computation~Program analysis}
\ccsdesc[500]{Software and its engineering~Formal software verification}
\ccsdesc[300]{Software and its engineering~Automated static analysis}


%%Bring back in camera-ready version
% \keywords{Abstract interpretation, Deductive verification, Large language models, Minimal contract, Undefined behaviors}

\maketitle

\setlength{\floatsep}{.8\baselineskip}
\setlength{\textfloatsep}{.8\baselineskip}
\setlength{\intextsep}{.8\baselineskip}
\captionsetup{aboveskip=5pt, belowskip=0pt}

\section{Introduction}

Runtime errors (RTEs), such as division by zero, buffer/numeric overflows, and null pointer dereference, are a common cause of \emph{undefined behaviors} (UBs) exhibited during the execution of C/C++ programs~\cite[Sect. 3.5.3]{c-standard}. These UBs can trigger catastrophic failures, rendering them critical considerations for safety-critical applications~\cite{ariane-5,buffer-overflow,resource-leak}.
%such as aerospace software~\cite{ariane-5} and medical devices~\cite{RTE-medical-devices,medical-devices}.\cmscomment{\atwzy Rephrase.} 
Consequently, in conventional program verification methodologies, establishing \emph{RTE-freeness} (i.e., conformance to the C standard specification) constitutes a necessary precondition for verifying \emph{functional correctness} (i.e., adherence to intended behaviors)~\cite{fm09-airbus}. State-of-the-art static analyzers based on \emph{abstract interpretation} \cite{cousotAbstractInterpretationUnified1977}, such as Astr{\'e}e~\cite{absint.astree} and \eva~\cite{eva-plugin}, %, and \mopsa~\cite{mopsa}
aim to reliably detect all potential UBs in large-scale C programs, thereby formally certifying the absence of RTEs. However, due to the inherent abstraction mechanism that soundly approximates concrete program semantics~\cite{cousotAbstractInterpretationUnified1977}, these tools often emit numerous false positives. Manually identifying such false alarms or tuning analyzer configurations for better accuracy remains notoriously difficult~\cite{parf,edf}.

% \blue{Specification synthesis... Refer to AutoSpec, X509-parser, \#Spec to briefly introduce specification synthesis (and relationships with program verification) and why and how we can use ss for proving the absence of RTEs. (See the Section 2 Backgrounds.) Then give the main challenges of current ss methods.}

%Program verification tools employing \emph{(automated) deductive verification}~\cite{deductive-verification}
Program verification tools, such as \emph{(automated) deductive verifiers}~\cite{deductive-verification},
%\wzycomment{Different from interactive proof assistants. Do not explicitly exclude BMC, which also requires loop invariants.}, 
provide a rigorous means for ensuring critical properties, including RTE-freeness and functional correctness. The verification process typically involves two stages: (i) constructing \emph{specifications} to formalize intended program behaviors, and (ii) proving that the program adheres to the specifications. %\wzycomment{Cousot formally proved that the former is harder than the latter.\cite[Sect. 1.3]{2018-research-progress-on-SA}}
While modern verifiers such as \Wp~\cite{frama-c/wp} and Dafny~\cite{dafny} automate the latter stage, the former -- known as \emph{specification synthesis}~\cite{specification-synthesis} -- still relies heavily on human expertise, especially for large-scale programs.~\cite{x509-parser,fm21-jcvm-verification}. 

% The communities of formal methods, software engineering, programming languages, and artificial intelligence have witnessed a recent surge of interest in leveraging large language models (LLMs) for \emph{automated specification synthesis}, demonstrating substantial improvements over conventional techniques.

% \blue{Use a paragraph to illustrate that most recent works do not deal with RTE-freeness verification, which is a combination of static analysis and deductive verification. See \cref{sec:potential-RTE-Guided-Verif} for more details.}

Recent studies~\cite{cav24-autospec,icse25-specgen,ASE24-Wu,ASE24-Pirzada,ICLR24-lemur} have explored large language models (LLMs) for \emph{automated specification synthesis}, demonstrating substantial improvements over conventional techniques. Nevertheless, these methods exhibit significant \emph{scalability limitations} when applied to real-world programs with, e.g., thousands of lines of code (LoC). %Current evaluations remain confined to small-scale benchmarks (e.g., SV-COMP test suites~\cite{sv-comp24}), each comprising individual or just a few functions.
% The reasons are primarily twofold: First, the \emph{context-length limitation} of LLMs~\cite{LLM-context-length-limitation} prohibits them from processing large-scale programs as a whole; Second, verifying large-scale programs -- which typically feature complex call hierarchies -- necessitates the synthesis of \emph{a diverse set of interprocedural specifications}, e.g., \emph{contracts} (pre- and postconditions), loop invariants, etc. The latter is beyond the capability of existing approaches: (i) some focus exclusively on intraprocedural specifications of specific categories, e.g., invariants~\cite{ICLR24-lemur, ASE24-Pirzada, ASE24-Wu} and assertions~\cite{oopsla25-LLM4assertion}, and (ii) others generate interprocedural specifications holistically without modeling intrinsic differences between preconditions and postconditions~\cite{cav24-autospec,icse25-specgen}, which are critical for identifying or validating the absence of RTEs (as demonstrated in \cref{sec:motivation}). A more detailed review of related work is provided in \cref{sec:related-work}.
The reasons are primarily twofold. First, the \emph{long-context reasoning limitation} of LLMs prohibits them from processing large-scale programs as a whole: (i) the entire program may exceed the maximum context window length~\cite{LLM-context-length-limitation}; and (ii) even when the window is sufficiently long, LLMs often struggle to effectively locate and utilize the precise information needed to solve the task in one shot~\cite{long-context-reasoning-1,long-context-reasoning-2}. Second, verifying large-scale programs -- which typically feature complex call hierarchies -- necessitates the synthesis of \emph{a diverse set of interprocedural specifications}, e.g., \emph{contracts} (pre- and postconditions), loop invariants, etc. The latter is beyond the capability of existing approaches: (i) some focus exclusively on intraprocedural specifications of specific categories, e.g., invariants~\cite{ICLR24-lemur, ASE24-Pirzada, ASE24-Wu} and assertions~\cite{oopsla25-LLM4assertion}, and (ii) others generate interprocedural specifications holistically without modeling intrinsic differences between preconditions and postconditions~\cite{cav24-autospec,icse25-specgen}, which are critical for identifying or validating the absence of RTEs (as demonstrated in \cref{sec:motivation}). A more detailed review of related work is provided in \cref{sec:related-work}.

In response to the aforementioned challenges, we present {\ours} -- \emph{an LLM-assisted framework for synthesizing fine-grained formal specifications that enable the effective verification of large-scale programs.} \ours adopts a \emph{divide-and-conquer} strategy orchestrating two synergistic phases: Phase 1 (\emph{Divide}) decomposes the monolithic task of RTE-freeness verification into manageable, prioritized units, guided by RTE assertions emitted by a static analyzer; Phase 2 (\emph{Conquer}) employs an LLM to infer interprocedural specifications with fine-grained strategies tailored to different program contexts pertaining to the units. We show that %{\ours}
this divide-and-conquer methodology facilitates (i) the \emph{synergy between static analysis and deductive verification}: The RTE assertions reported by static analysis are used to either construct necessary specifications certifying RTE-freeness or help locate root causes triggering RTEs (\`a la the \emph{minimal contract} paradigm \cite{minimal-contract}); and (ii) \emph{modular synthesis of interprocedural specifications} (at the granularity of verification units) and thereby a viable approach to the automated verification of large-scale programs.

We have implemented \ours on top of the deductive verifier \Wp and abstract interpretation-based analyzers \eva and \rte~\cite{frama-c/rte}.
% and compared it against the state-of-the-art LLM-based specification synthesis method \autospec and its enhanced variant \autospecRte over Frama-C-Problems benchmark. 
Experimental results show that \ours substantially outperforms state-of-the-art LLM-based approaches in terms of the rate of successfully verified benchmarks. In particular, \emph{\ours achieves highly automated RTE-freeness verification for real-world programs with over a thousand LoC, with a reduction of 80.6\%$\sim$88.9\% human verification effort} (measured by the number of human-intervened specifications). Moreover, {\ours} facilitates the identification of 6 confirmed RTEs in a practical spacecraft control system (1,280 LoC, 48 functions). These results demonstrate \ours's effectiveness in synthesizing high-quality specifications and scalability to large projects.

\vspace*{-.5mm}
\paragraph{\bf Contributions} The main contributions of this paper are summarized as follows.
\begin{itemize}
    \item We propose and formalize the concept of \emph{potential RTE-guided specification synthesis} as an effective means for LLM-aided specification synthesis towards scalable program verification.
    \item We present {\ours} -- \emph{a modular, fine-grained framework for inferring formal specifications} that integrates static analysis and deductive verification in a sound and terminating manner. \ours is, to the best of our knowledge, the first automated method capable of proving RTE-freeness of real-world programs with over $1,000$ LoC (with marginal human effort).
    \item We construct an open-source dataset consisting of real-world C programs annotated with specifications generated by \ours and crafted by experts for RTE-freeness verification. 
    \item We implement \ours and demonstrate its effectiveness and scalability on the benchmark dataset. We show that {\ours} outperforms state-of-the-art LLM-based approaches, and achieves highly automated RTE-freeness verification of large-scale, real-world programs.
\end{itemize}

\vspace*{-1mm}

% \paragraph{Paper Structure}
\paragraph{Paper Structure}
\cref{sec:preliminaries} recaps preliminaries on program verification. \cref{sec:motivation} motivates our conceptual idea of potential RTE-guided verification. \cref{sec:problem-statement} formalizes the problem of potential RTE-guided specification synthesis. \cref{sec:methodology} presents the technical underpinnings of our \ours framework. \cref{sec:evaluation} reports the evaluation results. \cref{sec:limitations} addresses several limitations and future directions. A review of related work is given in \cref{sec:related-work}. The paper is concluded in \cref{sec:conclusion}.
%Detailed proofs \blue{and other additional materials}\emph{Remove this?} are found in the appendix.
\section{Preliminaries}\label{sec:preliminaries}

Program verification seeks to determine whether a program conforms to a given formal specification. This section recaps several basic concepts in program verification. The discussion is not confined to specific programming or specification languages; rather, we focus on general imperative programs (with instruction-based execution) assessed against predicate-based specifications.

% Consider a fixed imperative program with variables $\texttt{vars}$ taking values in $\texttt{vals}$:

\paragraph{Program States and Configurations}
    A \emph{program state} $\pstate$ maps every program variable in $\vars$ to its value in $\vals$. We denote the set of program states by $\States \defeq \{\pstate \mid \pstate\colon \vars \to \vals\}$. A \emph{program configuration} $c \defeq \langle \pstate, \iota \rangle$ pairs a state $\pstate \in \States$ with the next instruction $\iota$ to be executed, representing the program's instantaneous state immediately before executing $\iota$.

\begin{example}\label{ex1-state}
    Consider a program snippet in C: \enquote{\texttt{\textbf{int} x = 1; x++;}}. The configuration $\langle \{\texttt{x} \mapsto 1\}, \texttt{x++} \rangle$ captures the program state where variable \texttt{x} has value 1 immediately before executing \texttt{x++}.
    \qedT
\end{example}

% \begin{definition}[Assertion]
%     Both are properties $q = \langle l, \phi \rangle$ where the verifier must establish $\phi$'s validity at $l$. When $\phi$ holds, execution proceeds without any change in both cases. Otherwise, if $\phi$ can not be proven, the execution diverges:
%     \begin{itemize}
%         \item \emph{Assumption}: Continues exclusively with program states satisfying $\phi$;
%         \item \emph{Assertion}: Halts and emits a violation alert.
%     \end{itemize}
% \end{definition}

\paragraph{Properties}
    A \emph{property} is a tuple $p \defeq \langle \phi, \iota \rangle$, where $\phi$ is a first-order logic \emph{predicate} (possibly with quantifiers) representing a subset of program states in $\States$ immediately before executing instruction $\iota$. We say that state $\pstate$ satisfies predicate $\phi$, written as $\pstate \models \phi$, iff $\pstate$ is in the set specified by $\phi$, i.e., $\phi(\pstate) = \true$; and $\pstate \not\models \phi$ otherwise. Moreover, a configuration $c = \langle \pstate, \iota' \rangle$ satisfies property $p  = \langle \phi, \iota \rangle$, denoted by $c \models p$, iff $\pstate \models \phi$ and $\iota' = \iota$; otherwise, it is denoted by $c \not\models p$.

In this paper, the terms \emph{specifications}, \emph{hypotheses}, and \emph{assertions} all refer to instances of properties.

% \begin{example}
%     \cref{fig-1:example-analysis-verification}~(b) illustrates an assertion $p_1 = \langle \phi_1, \iota_1 \rangle$ generated by the static analyzer \eva for the program \texttt{abs.c} depicted in \cref{fig-1:example-analysis-verification}~(a). Specifically, the predicate $\phi_1 = \red{\texttt{INT\_MIN < x}}$ encodes the necessary condition to prevent a potential integer overflow UB during the execution of $\iota_1$, as detailed in \cref{sec2-2:potential-RTE-Guided-Verif}.
% \end{example}

\begin{figure}[t]
	\centering
    \includegraphics[width=\linewidth]{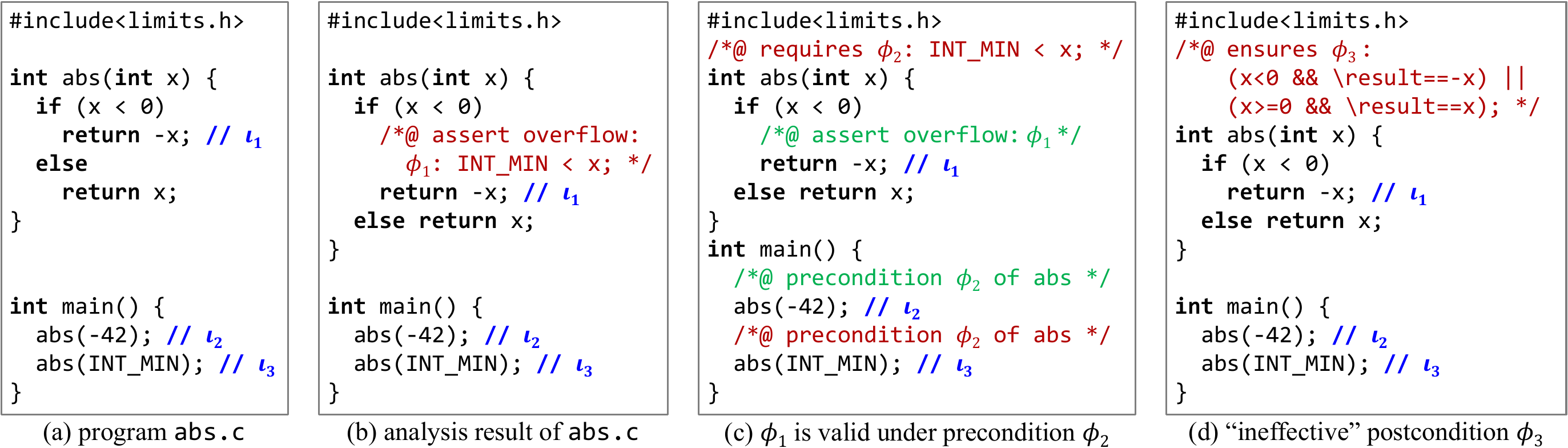}
	% \caption{Identifying potential RTEs in a C program via the abstract interpretation-based analyzer {\eva}~\cite{eva-plugin} and the deductive verifier {\Wp}~\cite{frama-c/wp}.}
    \caption{Potential RTE-guided verification process (a)(b)(c) v.s. unguided verification (a)(d).}
	\label{fig-1:example-analysis-verification}
\end{figure}

\begin{example}\label{ex2-property}
    Consider the program \texttt{abs.c} given in \cref{fig-1:example-analysis-verification}~(a), which defines a function \texttt{abs} for computing the absolute value of an integer input \texttt{x}. The \texttt{main} function calls \texttt{abs} twice, respectively with argument \texttt{-42} and \texttt{INT\_MIN}. A potential integer-overflow UB may occur when \texttt{x} evaluates to \texttt{INT\_MIN} immediately before executing instruction $\iota_1$ (the return statement). As depicted in \cref{fig-1:example-analysis-verification}~(b), an abstract interpretation-based static analyzer (e.g., \eva) detects this risk and inserts an assertion $p_1 = \langle \phi_1, \iota_1 \rangle$ at the potential UB point, using the ANSI/ISO C Specification Language (ACSL)~\cite{acsl}. The predicate $\phi_1 = \red{\texttt{INT\_MIN} < \texttt{x}}$ encodes the necessary constraint to prevent integer overflow during the execution of $\iota_1$, thereby serving as an RTE guard.
    \qedT
\end{example}

\begin{definition}[Trace and Subtrace]
    A \emph{trace} $\tau$ is a finite sequence of program configurations, i.e., 
    \begin{align}\label{eq:trace}
        \tau\, \ddefeq \left(c_0, \ldots, c_t\right) \eeq \left(\langle \sigma_0, \iota_0 \rangle, \ldots, \langle \sigma_t, \iota_t \rangle\right)~.
    \end{align}%
    Here, $\pstate_0$ is the initial state; and the state transition at each step $i \geq 0$ is governed by $\sigma_{i+1} = \llbracket \iota_i \rrbracket(\sigma_i)$, where $\llbracket \iota_i \rrbracket$ denotes the language-specific, single-step operational semantics of instruction $\iota_i$ \textnormal{\cite{MCCARTHY196333,floyd1967assigning}}. The \emph{length} of trace $\tau$ in \cref{eq:trace} is $t+1$. A trace $\tau' = (c'_0, \ldots, c'_s)$ is termed a \emph{subtrace} of $\tau$, denoted by $\tau' \prec \tau$, iff it forms a proper prefix of $\tau$, i.e., $s < t$ and $c'_i = c_i$ for any $0 \leq i \leq s$.
\end{definition}

\begin{example}
    The program fragment in \cref{ex1-state} admits, in principle, infinitely many traces since its initial state is unconstrained. One such trace is the sequence $\tau_1 = (\langle \texttt{x} \mapsto 0, \texttt{\textbf{int} x = 1} \rangle, \langle \{ \texttt{x} \mapsto 1 \}, \texttt{x++} \rangle, \langle \{ \texttt{x} \mapsto 2 \}, \iota_\text{null} \rangle)$, where $\iota_\text{null}$ denotes a null instruction signifying termination.
    \qedT
\end{example}

Next, we relate traces and properties via the notions of reachability and satisfaction:

\begin{definition}[Trace Reachability and Satisfaction]
    Given a property $p = \langle \phi, \iota \rangle$ and a trace $\tau = (c_0, \ldots, c_t) = (\langle \sigma_0, \iota_0 \rangle, \ldots, \langle \sigma_t, \iota_t \rangle)$, we say that $\tau$ \emph{reaches} $p$, denoted by $\tau \leadsto p$, iff $\iota_t = \iota$, i.e., the last instruction of $\tau$ coincides with $\iota$. Furthermore, we extend the satisfaction relation $\models$ to traces and properties: $\tau$ \emph{satisfies} $p$, written as $\tau \models p$, if and only if $\tau \leadsto p$ and $c_t \models p$.
    % Correspondingly, $\tau \not\models p \iff \neg(\tau \leadsto p) \lor c_t \not\models p$.
\end{definition}

% \begin{example}
%     \cref{fig-1:example-analysis-verification}~(b) depicts two traces, $\tau_2$ and $\tau_3$, starting at instructions $\iota_2$ and $\iota_3$ respectively, and both terminating at $\iota_1$ -- the instruction associated with assertion $p_1$ that constrains a potential integer overflow UB. Here, $\tau_2 \models p_1$ holds, whereas $\tau_3 \not\models p_1$ -- the argument \texttt{INT\_MIN} of instruction $\iota_3$ in trace $\tau_3$ triggers a genuine runtime error at $\iota_1$.
% \end{example}
\begin{example}\label{ex4:trace-reach-sat}
    Recall \cref{fig-1:example-analysis-verification}~(b) as discussed in \cref{ex2-property}. Consider two traces $\tau_2$ and $\tau_3$, which originate from instructions $\iota_2$ and $\iota_3$, respectively, and both terminate at $\iota_1$ -- the instruction associated with assertion $p_1$ that signifies a potential integer-overflow UB. We have $\tau_2 \leadsto p_1$, $\tau_3 \leadsto p_1$, $\tau_2 \models p_1$, yet $\tau_3 \not\models p_1$. The latter is due to the argument \texttt{INT\_MIN} of $\iota_3$ which triggers a genuine RTE at $\iota_1$.
    \qedT
\end{example}

The following definition captures interdependencies between specifications:

\begin{definition}[Property Validity under Hypotheses]\label{def:validity}
    Given two properties $p$ and $q$, we abuse the notation and write $p \models q$ to denote that $q$ is \emph{valid under hypothesis} $p$,
    %if and only if every trace $\tau$ that reaches $q$ and satisfies the following condition also satisfies $q$: for every subtrace $\tau'$ of $\tau$ that reaches $p$ (i.e., $\tau' \leadsto p$), it must hold that $\tau'$ satisfies $p$ (i.e., $\tau' \models p$). Formally, $p \models q$ holds 
    if and only if for every trace $\tau$ that reaches $q$, i.e., $\tau \leadsto q$, the following condition holds:
    \begin{align}\label{eq:validity}
        \underbrace{\forall \tau'\colon\, \left(\tau' \prec \tau \land \tau' \leadsto p \implies \tau' \models p\right)}_{\mathclap{\textnormal{every subtrace of $\tau$ that reaches $p$ also satisfies $p$}}} \qimplies \underbrace{\tau \models q}_{\mathclap{\textnormal{$\tau$ satisfies $q$}}}~.
    \end{align}%
    More generally, $q$ is \emph{valid under a set of hypotheses} $\mathcal{H}$, written as $\mathcal{H} \models q$, iff for any $\tau \leadsto q$,
    \begin{align}\label{eq:valid-under-hypo}
        \forall h \in \mathcal{H}.\ \forall \tau'\colon\, \left(\tau' \prec \tau \land \tau' \leadsto h \implies \tau' \models h\right) \qimplies \tau \models q~.
    \end{align}%
\end{definition}

% \begin{example}\label{ex5-validity}
%     \cref{fig-1:example-analysis-verification}~(c) illustrates a well-designed precondition $p_2 = \langle \phi_2, \iota_\text{pre}^\texttt{abs} \rangle$, where the predicate $\phi_2$ is identical to $\phi_1$ and $\iota_\text{pre}^\texttt{abs}$ denotes the first instruction following the entry point of function \texttt{abs}. This precondition validates the assertion $p_1$, and since $p_2 \models p_1$ holds, the function \texttt{abs} is certified RTE-free provided its precondition $p_2$ is guaranteed. Besides, by checking whether precondition $p_2$ is violated at the call sites (i.e., $\iota_2$ and $\iota_3$), we can distinguish the safe trace $\tau_2$ from the unsafe $\tau_3$ which triggers a genuine RTE (as detailed in \cref{ex4:trace-reach-sat}).
% \end{example}
\begin{example}\label{ex5-validity}
    \cref{fig-1:example-analysis-verification}~(c) demonstrates a well-designed precondition $p_2 = \langle \phi_2, \iota_\text{pre}^\texttt{abs} \rangle$ above function \texttt{abs}, where predicate $\phi_2$ is identical to $\phi_1$ and $\iota_\text{pre}^\texttt{abs}$ denotes the first instruction following the entry point of function \texttt{abs}.\footnote{Given a function \texttt{f}, we use $\iota_\text{pre}^\texttt{f}$ and $\iota_\text{post}^\texttt{f}$ to denote the first instruction after the entry point of \texttt{f} and a null instruction representing the termination of \texttt{f}, respectively. The call-site preconditions in \cref{fig-1:example-analysis-verification}~(c) are explained in \cref{sec:potential-RTE-Guided-Verif}.} This precondition ensures the validity of assertion $p_1$, namely $p_2 \models p_1$, and thus function \texttt{abs} is certified RTE-free under hypothesis $p_2$. An example of validity under \emph{multiple} hypotheses can be found in \cref{sec:feedback-driven-spec-syn}.
    %Furthermore, by verifying whether precondition $p_2$ is violated at call sites (i.e., $\iota_2$ and $\iota_3$), we can distinguish the safe trace $\tau_2$ from the unsafe trace $\tau_3$ that induces a genuine RTE (as elaborated in \cref{ex4:trace-reach-sat}).
    \qedT
\end{example}

Now, we have all the ingredients to formally describe a sound program verifier:

\begin{definition}[Sound Verifier]\label{def:verifier}
    Given a program $\textit{prog}$, a hypothesis set $\mathcal{H}$, and a target property $q$, a (deductive) \emph{verifier} is a function mapping these inputs to a certain verification result:
    \begin{align}\label{eq:verifier}
        \textit{Verify}\colon\ \left(\textit{prog}, \mathcal{H}, q\right)\! \mmmapsto \textit{status}~,
    \end{align}%
    where the output $\textit{status}$ is either (i) $\tagTrue$ signifying that $q$ is valid under $\mathcal{H}$; or (ii) $\tagUnknown$ indicating invalidity or verification failure due to resource exhaustion and/or solver limitations (e.g., employing over-approximations of program semantics rather than exact representations) \textnormal{\cite{decision-procedure}}.
    % and $\text{feedback}$ contains information such as proof obligations (a.k.a. verification conditions) of $q$. 
    % This two-valued output represents a sound over-approximation: the \red{\text{unknown}} status subsumes both the classical \texttt{false} (indicating invalidity) and cases where verification fails due to resource exhaustion, or solver limitations~\cite{decision-procedure}.
    In particular, the verifier in \cref{eq:verifier} is \emph{sound} if it satisfies the following soundness condition:
    % \begin{align*}
    %     \textit{Verify}(\textit{prog}, \mathcal{H}, q) = \green{\text{true}} &\implies \mathcal{H} \models q \\
    %     \textit{Verify}(\textit{prog}, \mathcal{H}, q) = \red{\text{unknown}} &\impliedby \mathcal{H} \not\models q
    % \end{align*}
    \begin{align}\label{eq:sound-verifier-1}
        \textit{Verify}\left(\textit{prog}, \mathcal{H}, q\right) \eeq \tagTrue &~\qimplies~ \mathcal{H} \mmodels q~.
        % \qand\\
        % \mathcal{H} \nnotmodels q &\qimplies
        % \textit{Verify}\left(\textit{prog}, \mathcal{H}, q\right) \eeq \tagUnknown~.
    \end{align}%
    This condition ensures that a $\tagTrue$ verdict guarantees the validity of $q$ under $\mathcal{H}$, while $\mathcal{H} \not\models q$ necessarily results in $\tagUnknown$. It then follows from the consequence rule of Floyd-Hoare Logic~\textnormal{\cite{hoare-logic}} that
    % \begin{equation}\label{eq:sound-verifier-2}
    %     \mathcal{H}\subseteq \mathcal{H}' \implies \big(\textit{Verify}(\textit{prog}, \mathcal{H}, q) = \green{\text{true}} \implies \textit{Verify}(\textit{prog}, \mathcal{H}', q) = \green{\text{true}}\big)
    % \end{equation}
    \begin{equation}\label{eq:sound-verifier-2}
        \textit{Verify}\left(\textit{prog}, \mathcal{H}, q\right) \eeq \tagTrue \aand \mathcal{H} \subseteq \mathcal{H}'
        \qimplies
        \textit{Verify}\left(\textit{prog}, \mathcal{H}', q\right) \eeq \tagTrue~,
    \end{equation}
    i.e., a property valid under hypotheses $\mathcal{H}$ must also be valid under logically stronger hypotheses $\mathcal{H}'$.
\end{definition}%
\noindent
The verifier-reported statuses ($\tagTrue$ or $\tagUnknown$) for all properties in \cref{fig-1:example-analysis-verification,fig-2:dependencies-spec-property,fig-3:gen-spec-based-on-feedback} are visually indicated using their corresponding colors. Moreover, in practice, a sound verifier can take an extended form
\begin{align}\label{eq:sound-verifier-extended}
    \textit{Verify}\colon\ \left(\textit{prog}, \mathcal{H}, q\right) \mmmapsto \left\langle \textit{status},\textit{feedback} \right\rangle~,
\end{align}%
which augments \cref{eq:verifier} with auxiliary \emph{feedback} information. Such feedback typically includes \emph{proof obligations} (aka, \emph{verification conditions}) -- logical formulas that, when being valid, suffice to guarantee program adherence to specifications. A status of $\tagTrue$ indicates that all proof obligations are valid, whereas a status of $\tagUnknown$ signifies that the verifier could neither prove nor refute the obligations. We show in \cref{sec:feedback-driven-spec-syn} that such feedback can be leveraged to refine specifications.

\section{Motivation}\label{sec:motivation}

\subsection{Potential RTE-Guided Verification}\label{sec:potential-RTE-Guided-Verif}

We use the simple example in \cref{fig-1:example-analysis-verification} to demonstrate our idea of \emph{potential RTE-guided verification}~\cite{preguss}, which aims to synergize between static analysis and deductive verification (in line with the \emph{minimal contract} paradigm \cite{minimal-contract}): We first apply a static analyzer to the source program in \cref{fig-1:example-analysis-verification}~(a); if any RTE alarm is reported in the form of an assertion, e.g., $p_1 = \langle \phi_1, \iota_1 \rangle$ in \cref{fig-1:example-analysis-verification}~(b) (as explained in \cref{ex2-property}), we construct necessary specifications, e.g., the precondition $p_2 = \langle \phi_2, \iota_\text{pre}^\texttt{abs} \rangle$ in \cref{fig-1:example-analysis-verification}~(c) (as elaborated in \cref{ex5-validity}), which assist a verifier in certifying the assertion. Both the assertion and the precondition then serve as \emph{guard assertions} to ensure RTE-freeness of the source program. \cref{fig-1:example-analysis-verification}~(c) demonstrates this mechanism: By verifying whether the (weakest) precondition $p_2$ is violated at call sites of function \texttt{abs} (i.e., $\iota_2$ and $\iota_3$), we can distinguish safe traces (e.g., $\tau_2$) from unsafe ones (e.g., $\tau_3$) that induce a \emph{definite} RTE.

\emph{Can LLMs generate such specifications?} Taking \texttt{abs.c} as an example, most advanced LLMs -- when fed with the source program context \emph{and} the RTE assertion $p_1$ -- exhibit strong code comprehension capabilities and can readily generate weakest preconditions semantically equivalent to $p_2$. However, state-of-the-art approaches to LLM-based specification synthesis~\cite{ICLR24-lemur,ASE24-Wu,ASE24-Pirzada,cav24-autospec,icse25-specgen} do not exploit RTE assertions emitted by established static analysis techniques (most pertinently, abstract interpretation \cite{cousotAbstractInterpretationUnified1977}), thus significantly impairing their performance in generating necessary specifications for RTE-freeness verification. For instance, it is common for LLMs to produce specifications such as the postcondition $p_3 = \langle \phi_3, \iota_\text{post}^\texttt{abs} \rangle$ depicted in \cref{fig-1:example-analysis-verification} (d), given \emph{only} the source program \texttt{abs.c} in \cref{fig-1:example-analysis-verification}~(a). The predicate $\phi_3=$ \red{\texttt{\detokenize{(x<0 && \result==-x)}}} \red{\texttt{\detokenize{||}}} \red{\texttt{\detokenize{(x>=0 && \result==x)}}} appears to correctly describe the behavior of \texttt{abs}, except for the case \texttt{x = INT\_MIN}, leading verifiers to mark it as $\tagUnknown$. Moreover, postcondition $p_3$ is \enquote{ineffective} for detecting the actual RTE source at $\iota_3$. Hence, we have

\begin{insight}
\label[insight]{ins:RTE-assertions}
One can leverage RTE assertions from abstract interpretation-based \mbox{analyzers} -- as targets for RTE-freeness verification -- to guide LLMs to synthesize specifications effectively.
\end{insight}

Then, the following question naturally arises: \emph{Does it suffice to extend existing LLM-based synthesis approaches by simply feeding the LLM therein with RTE assertions?} The answer is unfortunately negative: We discuss the corresponding technical challenges throughout the rest of this section and report the practical insufficiency of such a na{\" i}ve extension in \cref{sec:evaluation} (RQ1).

\subsection{Interprocedural Specifications}\label{sec:interprocedural-spec}

\begin{figure}[t]
	\centering
    \includegraphics[width=\linewidth]{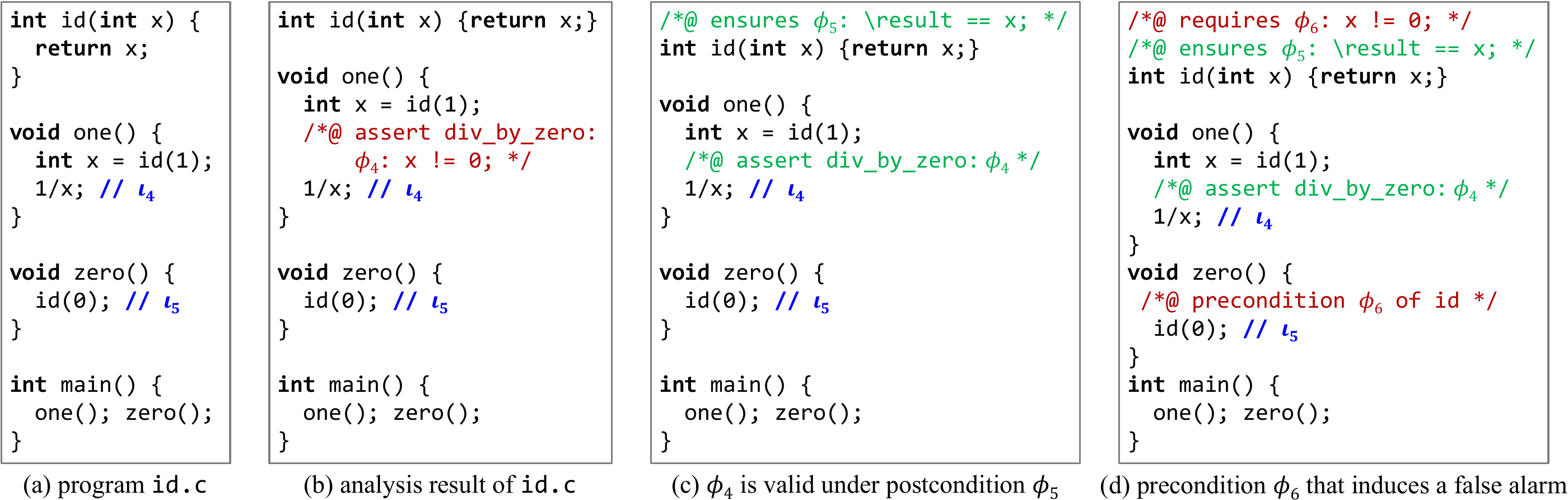}
	\caption{The dual role of interprocedural specifications: postconditions can help discharge false RTEs (c) while over-constrained preconditions can induce false alarms (d).}
	\label{fig-2:dependencies-spec-property}
\end{figure}

The example in \cref{fig-1:example-analysis-verification} illustrates the simple case where an RTE assertion (i.e., $p_1$) can be validated through specifications (i.e., $p_2$) \emph{localized to its host function} (\texttt{abs}). This approach with localized specifications is, however, insufficient for ensuring RTE-freeness in real-world programs featuring \emph{complex call hierarchies}. Consider the program \texttt{id.c} in \cref{fig-2:dependencies-spec-property}~(a), which defines a function \texttt{id} that returns the identical value of its input $x$, alongside functions \texttt{one} and \texttt{zero} that invoke \texttt{id} with arguments \texttt{1} and \texttt{0}, respectively. The \texttt{main} function calls both \texttt{one} and \texttt{zero}. Static analysis reveals a potential division-by-zero UB in function \texttt{one}, flagged by the assertion $p_4 = \langle \phi_4, \iota_4 \rangle$ in \cref{fig-2:dependencies-spec-property}~(b) (with $\phi_4 = \red{\texttt{x}}\ \,\red{\texttt{!=}} \ \,\red{\texttt{0}}$). Unlike the assertion $p_1$ in \cref{fig-1:example-analysis-verification}, constructing any precondition for \texttt{one} does not suffice to validate $p_4$, as \texttt{one} lacks parameters and thus imposes no constraints on its inputs.

To eliminate such false alarms (e.g., $p_4$), which frequently arise in practice, it is necessary to employ \emph{interprocedural specifications} -- specifications that extend beyond the host functions of target assertions. For instance, the postcondition $p_5 = \langle \phi_5, \iota_\text{post}^\texttt{id} \rangle$ for function \texttt{id} in \cref{fig-2:dependencies-spec-property}~(c) (with $\phi_5 =$ \green{\texttt{\detokenize{\result == x}}}) serves precisely this role. Specifically, the value of \texttt{x} at $\iota_4$ is determined by argument \texttt{1} and the postcondition of \texttt{id} at the call site \texttt{id(1)}, thereby validating $p_4$ , i.e., $p_5 \models p_4$.

The synthesis of interprocedural specifications, as exemplified by postcondition $p_5$, depends on LLMs comprehending interprocedural program contexts that span across multiple functions. Specifically, LLMs must realize that discharging assertion $p_4$ in function \texttt{one} necessitates constructing a postcondition for function \texttt{id}. Nonetheless, current approaches to interprocedural specification synthesis, e.g., \cite{icse25-specgen,cav24-autospec}, exhibit two key limitations: (i) They provide LLMs with the entire program as context, which is constrained by the \emph{finite context window} of LLMs and thus do not scale to large programs; (ii) Even for small programs with full context, LLMs may generate \emph{over-constrained preconditions} due to \emph{hallucination}~\cite{hallucination}, thus yielding spurious false alarms. As illustrated in \cref{fig-2:dependencies-spec-property}~(d), an LLM might be misled by the context ``\red{\texttt{x != 0}} ($\phi_4$) and \texttt{1/x}'' at $\iota_4$, and thus forge an over-constrained precondition $p_6 = \langle \phi_6, \iota_\text{pre}^\texttt{id} \rangle$ for \texttt{id} (with $\phi_6 =$ \red{\texttt{x != 0}}), which triggers a false alarm at the call site $\iota_5$ (\texttt{id(0)}). Although discarding over-constrained preconditions while retaining postconditions may resolve immediate false alarms (as in \cref{fig-2:dependencies-spec-property}~(c)), determining whether a precondition is over-constrained is per se a nontrivial task (cf. \cref{fig-1:example-analysis-verification}~(c) vs. \cref{fig-2:dependencies-spec-property}~(d))~\cite{necessary-preconditions-1,necessary-preconditions-2}: In \cref{fig-1:example-analysis-verification}~(c), indiscriminately discarding the precondition compromises soundness (as true RTEs are missed).
% and thus undermines verification integrity. Consequently, interprocedural specification synthesis mechanisms must prioritize preventing the generation of over-constrained preconditions.
In a nutshell, we observe
%
% \vspace*{-1mm}
\begin{challenge}
\label[challenge]{cha:LLM-context-length}
In the presence of long-context reasoning limitations, how to instruct LLMs to access particular program contexts required for generating interprocedural specifications for large-scale programs with complex call hierarchies?
\end{challenge}

\vspace*{-2mm}

\begin{challenge}
\label[challenge]{cha:fine-grained}
How to design a fine-grained interprocedural-specification synthesis mechanism -- e.g., determining which specifications to generate, at which program locations, and in what sequence -- to alleviate the synthesis of over-constrained preconditions\footnotemark?
\end{challenge}

\footnotetext{While over-constrained specifications beyond preconditions (e.g., postconditions and loop invariants) can also lead to issues such as false negatives, these can be addressed through simpler mechanisms, as discussed in \cref{sec4-2-3:syntax-semantics-check}.}

\subsection{Feedback-Driven Specification Refinement}\label{sec:feedback-driven-spec-syn}

\begin{figure}[t]
	\centering
    \includegraphics[width=\linewidth]{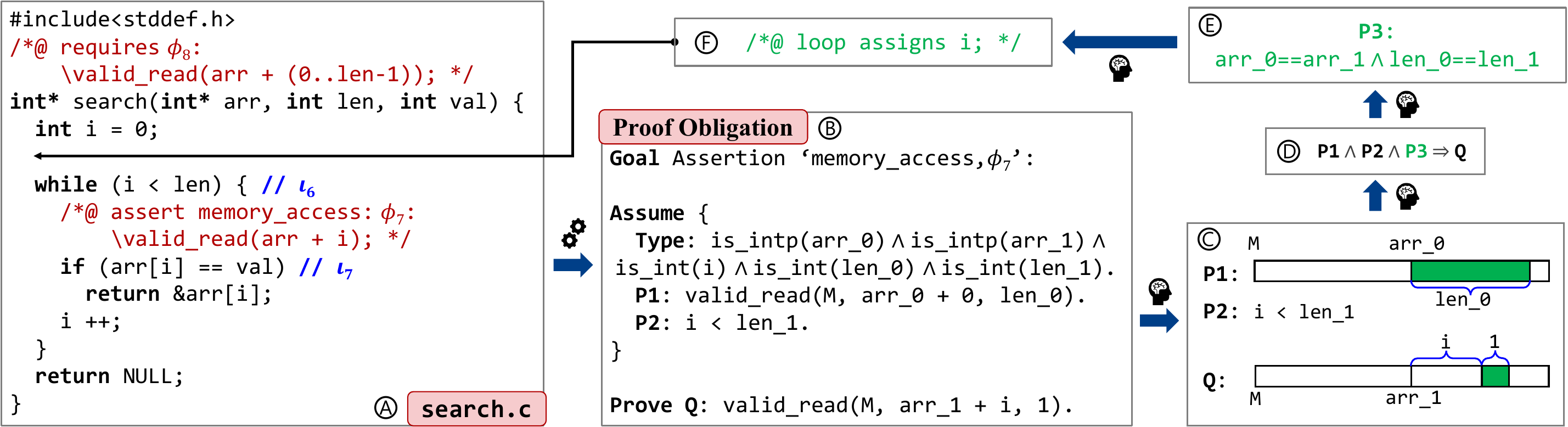}
	\caption{Handcrafting missing specifications via verifier feedback (e.g., proof obligations) for validating $p_7$.}
	\label{fig-3:gen-spec-based-on-feedback}
\end{figure}

\cref{sec:interprocedural-spec} has demonstrated the challenges in constructing interprocedural specifications; however, validating RTE assertions for individual functions can also be difficult. Consider, for example, the program \texttt{search.c} in \cref{fig-3:gen-spec-based-on-feedback}~\Circled{A}, which defines a function \texttt{search} that iterates through an integer array \texttt{arr} of length \texttt{len} and returns either the address of the first element matching the target value \texttt{val}, or \texttt{NULL} if no such element exists. Static analysis reveals a potential invalid-memory-access UB, flagged by the assertion $p_7 = \langle \phi_7,\iota_7\rangle$ in \Circled{A} (where $\phi_7 =$ \red{\texttt{\string\valid\_read(arr + i)}}. Guided by $p_7$, LLMs can generate a precondition $p_8 = \langle\phi_8, \iota_\text{pre}^\texttt{search}\rangle$, where $\phi_8 =$ \red{{\texttt{\string\valid\_read(arr + (0..len-1))}}} asserts read validity for the entire array \texttt{arr} with length \texttt{len}. Counterintuitively, the verifier fails to certify the validity of accessing \texttt{a[i]} (i.e., marking $p_7$ at $\iota_7$ as $\tagUnknown$) under the hypothesis $p_8$.

This behavior stems from the fact that modern deductive verifiers employ elaborate mechanisms to handle complex C semantics (e.g., memory access at $\iota_7$ in \Circled{A}). This necessitates \emph{understanding the verifier's internal mechanisms and verification details} when validating RTE assertions involving such semantic features. \Cref{fig-3:gen-spec-based-on-feedback}~\Circled{B}-\Circled{F} demonstrates a step-by-step process by a verification expert to construct missing specifications for discharging $p_7$ leveraging \emph{verifier feedback}, including the proof obligation of $p_7$ in \Circled{B}.
Specifically, the obligation comprises hypotheses \texttt{\textbf{P1}} (precondition $p_8$), \texttt{\textbf{P2}} (loop condition \texttt{i} < \texttt{len} at $\iota_6$), and type declarations of integer pointers (\texttt{arr\_0}, \texttt{arr\_1}) and integer variables (\texttt{i}, \texttt{len\_0}, \texttt{len\_1}), associated with the goal \texttt{\textbf{Q}} to be verified.
Here, \texttt{\textbf{Q}} is a transformed version of $p_7$ that is used internally by \Wp. In \texttt{\textbf{Q}}, \texttt{M} denotes a simplified \enquote{typed} memory model (for integers in this case), as outlined in~\cite[Sect. 4.3.4]{frama-c-verification}, where each memory block has the size of an \texttt{int} (i.e., 4 bytes), forming a region storing integer arrays such as \texttt{arr\_0} and \texttt{arr\_1}. The variable \texttt{arr\_1} represents the base address of array \texttt{arr} at $\iota_7$. It is distinguished from \texttt{arr\_0} -- the base address of \texttt{arr} in precondition $p_8$ -- because \texttt{arr} is mutable and \Wp assumes by default that it may be modified inside loops.
At this stage, the assertion $p_7$ remains unproven because $\texttt{\textbf{P1}} \land \texttt{\textbf{P2}} \Rightarrow \texttt{\textbf{Q}}$ cannot be established by external solvers.

The verification expert can model the semantics of \texttt{\textbf{P1}} and \texttt{\textbf{Q}} through memory representations as depicted in \Circled{C}. \texttt{\textbf{P1}}: \texttt{valid\_read(M,arr\_0+0,len\_0)} denotes read validity for a memory block starting at \texttt{arr\_0} spanning \texttt{len\_0} elements, whilst \texttt{\textbf{Q}}: \texttt{valid\_rd(M,arr\_1+i,1)} asserts read validity for a single element at \texttt{arr\_1+i}. Although no solver proves $\texttt{\textbf{P1}} \land \texttt{\textbf{P2}} \Rightarrow \texttt{\textbf{Q}}$, experienced practitioners may strengthen the \emph{antecedent} by conjoining \green{\textbf{\texttt{P3}}} and evaluating $\texttt{\textbf{P1}} \land \texttt{\textbf{P2}} \land \green{\texttt{\textbf{P3}}} \Rightarrow \texttt{\textbf{Q}}$ (see \Circled{D}). The missing predicate \green{\textbf{\texttt{P3}}} is immediately derived (based on verification expertise) as \green{\texttt{arr\_0 == arr\_1} $\land$ \texttt{len\_0 == len\_1}} (\Circled{E}). This reveals the verifier's over-approximation of side effects -- assuming that variables defined outside the loop, such as array address \texttt{arr} and length \texttt{len}, could be modified inside the loop scope. Consequently, adding the specification \green{\texttt{loop assigns i}} (denoted by $p_9$) becomes essential to restrict reassignment exclusively to variable \texttt{i} (\Circled{F}). Inserting this specification above the loop condition ($\iota_6$) enables verifiers to eliminate the false invalid-memory-access alarm, i.e., $\{p_8, p_9\} \models p_7$ as per \cref{eq:valid-under-hypo}. This gives rise to the idea of \emph{feedback-driven specification refinement}:

% The manual specification synthesis process motivates leveraging verifier feedback -- beyond mere program context -- when prompting LLMs. Given LLMs' susceptibility to \emph{hallucination}~\cite{hallucination}, they frequently generate specifications exhibiting critical flaws: (i) \emph{syntactically illegal} (violating specification language grammar) or (ii) \emph{semantically unsatisfiable} (contradicting program behaviors). Verifier feedback -- particularly proof obligations and error logs -- can be strategically incorporated to prompt LLMs to either correct erroneous specifications or regenerate valid candidates, thereby transforming failed verification attempts into refinement opportunities.
\vspace*{-1.3mm}

\begin{insight}
\label[insight]{ins:verifier-feedback}
The feedback from a verifier, in particular, the proof obligations that remain to be discharged, can be leveraged to refine the specifications generated by LLMs effectively.
\end{insight}

\section{Problem Formulation}\label{sec:problem-statement}

This section formalizes the problem of potential RTE-guided specification synthesis. To this end, we first introduce a formal characterization of RTE-freeness:

\begin{definition}[RTE-Freeness]
    Given a program $\textit{prog}$, a set of RTE assertions $\mathcal{A}$, and a set of specifications $\mathcal{S}$, we say that $\textit{prog}$ is \emph{free-of-RTE} if and only if
    \begin{equation}\label{eq:free-RTE}
        \forall p \in \mathcal{A} \cup \mathcal{S}\colon\ \mathcal{A} \cup \mathcal{S} \setminus \{p\} \mmodels p~,
    \end{equation}%
    namely, the properties in $\mathcal{A} \cup \mathcal{S}$ are mutually consistent -- every property in the set is valid under the rest. More generally, $\textit{prog}$ is said to be \emph{free-of-RTE under hypotheses $\mathcal{H} \subseteq \mathcal{A}\cup\mathcal{S}$} if and only if
    \begin{equation}\label{eq:free-RTE-hypo}
        \forall p \in \mathcal{A}\cup\mathcal{S}\setminus\mathcal{H}\colon\ \mathcal{A}\cup\mathcal{S}\setminus\{p\} \mmodels p~.
    \end{equation}
    % \begin{equation}\label{eq:free-RTE-hypo}
    %     \exists \mathcal{H} \subseteq \mathcal{A}\cup\mathcal{S}.\ \forall p \in \mathcal{A}\cup\mathcal{S}\setminus\mathcal{H}\colon\ \mathcal{A}\cup\mathcal{S}\setminus\{p\} \mmodels p~.
    % \end{equation}
    Note that \cref{eq:free-RTE} is a special case of \cref{eq:free-RTE-hypo} with $\mathcal{H}=\emptyset$.
\end{definition}

% It follows that the characterization \cref{eq:free-RTE-hypo} boils down to \cref{eq:free-RTE} if the hypotheses are validated:
It follows that the characterization \cref{eq:free-RTE-hypo} boils down to \cref{eq:free-RTE} if the hypotheses $\mathcal{H}$ are discharged:

\begin{restatable}[RTE-Freeness under Valid Hypotheses]{theorem}{restateRTEFreeness}\label{theorem:RTE-freeness}
% \begin{theorem}[RTE-Freeness under Valid Hypotheses]\label{theorem:RTE-freeness}
    Suppose $\textit{prog}$ is free-of-RTE under hypotheses \(\mathcal{H}\) as per \cref{eq:free-RTE-hypo}. If every \(h \in \mathcal{H}\) is valid, i.e., $\models h$, then $\textit{prog}$ is free-of-RTE in the sense \mbox{of \cref{eq:free-RTE}}.
% \end{theorem}
\end{restatable}
%A formal proof of \cref{theorem:RTE-freeness} is provided in \cref{sec:appendix-rte-freeness}.

\begin{example}
    Recall the three real-world scenarios of RTE-freeness verification in \cref{fig-1:example-analysis-verification,fig-2:dependencies-spec-property,fig-3:gen-spec-based-on-feedback}:

    The program \texttt{abs.c} in \cref{fig-1:example-analysis-verification} is free-of-RTE under hypothesis $\{p_2\}$. The precondition $p_2$ facilitates detecting a real RTE in trace $\tau_3$ while safeguarding trace $\tau_2$. With $p_2$, one can correct the program by removing the instruction \texttt{abs(INT\_MIN)} ($\iota_3$) or revising the implementation of \texttt{abs}. This exemplifies a classic \emph{source-to-sink} problem~\cite{source-and-sink} in static (taint) analysis, where the goal is to identify that \texttt{abs(INT\_MIN)} at $\iota_3$ (source) triggers an overflow RTE at $\iota_1$ (sink). Potential RTE-guided specification synthesis provides a deductive verification mechanism for solving such problems.

    The program \texttt{id.c} in \cref{fig-2:dependencies-spec-property} is free-of-RTE. The postcondition $p_5$ suffices to establish $\models p_5 \models p_4$. This represents the ideal case for RTE-freeness verification, where all alarms raised by static analyzers are false positives and can be eliminated by deductive verifiers with appropriate specifications.

    The program \texttt{search.c} in \cref{fig-3:gen-spec-based-on-feedback} is free-of-RTE under hypotheses $\{p_8, p_9\}$. Unlike \texttt{abs.c}, precondition $p_8$ cannot be used to locate RTEs directly, as no caller function to \texttt{search} exists in the program. This case depicts the scenario of constructing \emph{summaries} for modules in large projects, known as \emph{modular static program analysis}~\cite{modular-analysis}. Preconditions like $p_8$ provide assurances that modules such as \texttt{search} are free-of-RTE if all callers adhere to these preconditions.
    \qedT
\end{example}

The \emph{potential RTE-guided specification synthesis problem} concerned in this paper reads as follows:

% \begin{tcolorbox}[boxrule=1pt,colback=white,colframe=black!75,boxsep=0mm]%[boxrule=0pt,colframe=black!75]
% \textbf{Problem Statement.}\ \ Given a program $\textit{prog}$ with a set of RTE assertions $\mathcal{A}$ annotated by an abstract interpretation-based static analyzer, a deductive verifier $\textit{Verify}$, and a large language model $\textit{LLM}$, automatically generate a set of specifications $\mathcal{S}$ and prove that $\textit{prog}$ is either (i) free-of-RTE, or (ii) free-of-RTE under hypotheses $\mathcal{H} \subseteq \mathcal{A}\cup\mathcal{S}$.
% \end{tcolorbox}

\begin{tcolorbox}[boxrule=1pt,colback=white,colframe=black!75,boxsep=0mm]%[boxrule=0pt,colframe=black!75]
\textbf{Problem Statement.}\ \ Given a program $\textit{prog}$ with a set of RTE assertions $\mathcal{A}$ annotated by an abstract interpretation-based static analyzer, %a deductive verifier $\textit{Verify}$, and a large language model $\textit{LLM}$, 
automatically generate a set of specifications $\mathcal{S}$ by leveraging a large language model $\textit{LLM}$, and prove with a deductive verifier $\textit{Verify}$ that $\textit{prog}$ is either (i) free-of-RTE, or (ii) free-of-RTE under hypotheses $\mathcal{H} \subseteq \mathcal{A}\cup\mathcal{S}$.
\end{tcolorbox}

\noindent
Note that case (ii) predominates in the verification of large-scale, real-world programs. The presence of a hypothesis \(h \in \mathcal{H}\) typically stems from one of the four scenarios: 
(a) \(h\) signifies a genuine RTE, which inherently cannot -- and should not -- be validated by any sound verifier; 
(b) \(h\) is a false alarm that fails to be discharged due to resource exhaustion or inherent solver limitations; 
(c) \(h\) is a false alarm resulting from an over-constrained precondition generated by the approach; or 
(d) \(h\) is a false alarm, but the specification-synthesis framework does not suffice to eliminate it -- for instance, generating precise specifications for behaviors involving complex data structures and/or intricate memory models remains notoriously hard; see, e.g., \cite{fm21-jcvm-verification} for dedicated research.

A practical strategy for addressing these scenarios involves manual validation of the hypotheses in \(\mathcal{H}\) (as justified by \cref{theorem:RTE-freeness}), supplemented by revising generated specifications (particularly for scenario (c)) and constructing additional specifications (particularly for scenario (d)). This approach proves substantially more feasible than attempting to verify RTE-freeness against the entire set of RTE assertions \(\mathcal{A}\) produced by the static analyzer, as demonstrated in \cref{sec:evaluation} (RQ2).

\section{Methodology}
\label{sec:methodology}

% This section presents the design principles behind {\ours} -- our framework for Potential Runtime Error-GUided Specification Synthesis. As depicted in \cref{fig-4:framework}, {\ours} is comprised of two synergistic phases. Specifically, Phase 1 (\cref{sec3-1:construction-prioritization}) employs a divide-and-conquer strategy~\cite{divide-and-conquer} for decomposing the monolithic RTE-freeness verification into prioritized units, each of which contains a target assertion to be validated together with necessary program context. Phase 2 (\cref{sec3-2:interprocedural-spec-syn}) infers necessary specifications via LLM agents along caller-callee chains to prove the target assertion per verification unit. Below we show how these two phases cooperate to verify the RTE-freeness for real-world programs with complex call hierarchies through the motivating example which has been illustrated in \cref{sec:interprocedural-spec} and \cref{fig-2:dependencies-spec-property}. % Below, we show how these two components cooperate to facilitate {\ours}'s scalability to large-scale programs.

This section presents {\ours} -- our framework for Potential Runtime Error-GUided Specification Synthesis. As sketched in \cref{fig-4:framework}, \ours adopts a \emph{divide-and-conquer} strategy (in a fashion similar to \cite{divide-and-conquer}) which operates through two synergistic phases: Phase 1 (\emph{Divide}) decomposes the monolithic task of RTE-freeness verification into manageable, prioritized units. Each unit contains a guard assertion together with its necessary program context for validation; Phase 2 (\emph{Conquer}) employs an LLM to infer interprocedural specifications along caller-callee chains, thereby certifying the target assertion in each verification unit. Below, we show how these two phases cooperate to enable RTE-freeness verification of real-world programs with complex call hierarchies.
%building upon the motivating examples introduced in \cref{sec:interprocedural-spec} and \cref{fig-2:dependencies-spec-property}.

% \subsection{Decomposing the Monolithic Verification}
\subsection{Phase 1: Potential RTE-Guided Construction and Prioritization of Verification Units}
\label{sec3-1:construction-prioritization}

\begin{figure*}
  \centering
  \includegraphics[width=\linewidth]{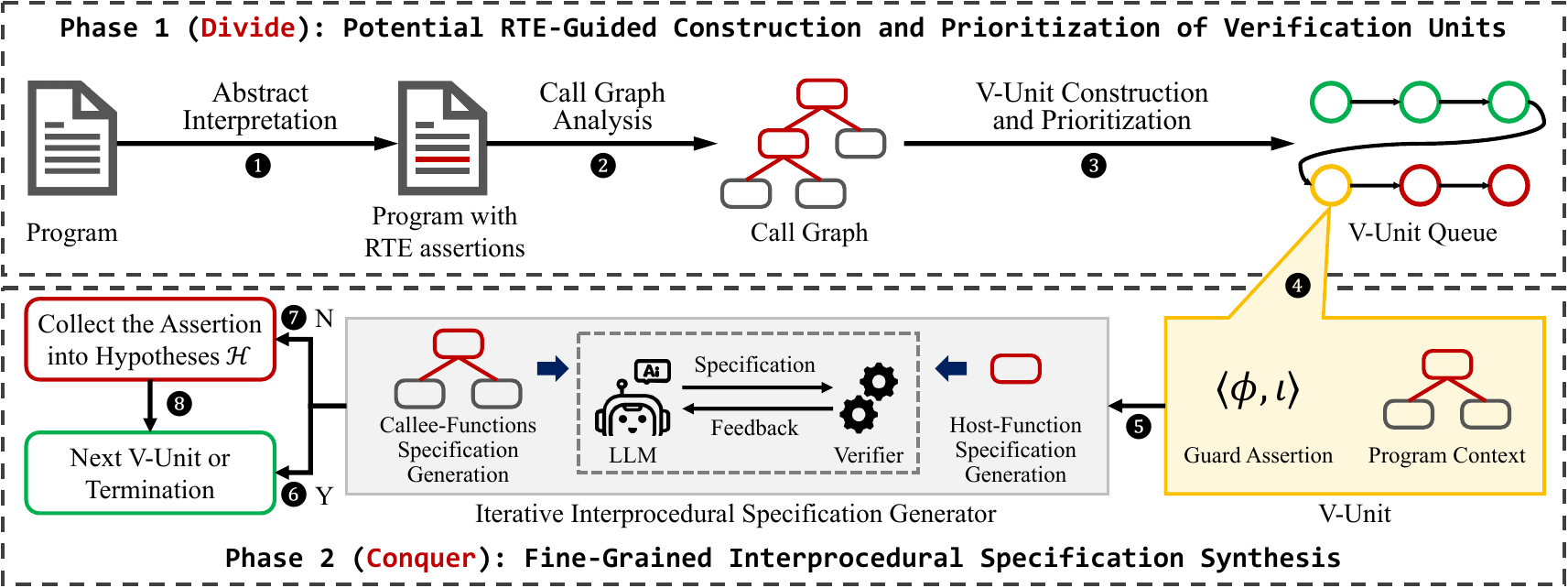}
  \caption{Divide-and-conquer architecture of the {\ours} framework.}
  \label{fig-4:framework}
\end{figure*}

To address \cref{cha:LLM-context-length} (long-context reasoning limitation of LLMs), Phase 1 decomposes the monolithic RTE-freeness verification task into a sequence of manageable \emph{verification units} (V-Units), as illustrated in \cref{fig-4:framework} (upper part). The process begins by employing abstract interpretation conducted by static analyzers (e.g., {\rte}~\cite{frama-c/rte} and {\eva}) to generate RTE assertions signifying \emph{all} potential UBs in the source program (\ding{182}). Subsequently, the program's \emph{call graph} is constructed while recording all of its call sites (\ding{183}), yielding a graph structure with a comprehensive set of \emph{guard assertions}. For each assertion, a \emph{V-Unit} is created, encapsulating both the assertion and its essential contextual program slices. These V-Units are then prioritized into a queue to streamline the verification workflow (\ding{184}). Below, we first elaborate on the generation of guard assertions, which serve as sub-goals for the holistic RTE-freeness verification (\cref{sec3-1-1:rte-guard-assertion}) and then introduce the V-Unit structure along with prioritization principles (\cref{sec3-1-2:v-unit}).

\subsubsection{Guard Assertion Generation}\label{sec3-1-1:rte-guard-assertion}

As discussed in \cref{sec:potential-RTE-Guided-Verif}, both the RTE assertions emitted by static analyzers and the synthesized preconditions act as \emph{guard assertions} to ensure RTE-freeness of the source program. Whereas RTE assertions can be generated via abstract interpretation (\ding{182}), preconditions must be constructed on-the-fly during specification synthesis. However, we observe that validating a precondition ($p=\langle \phi,\iota_\text{pre}^\texttt{f} \rangle$) involves two key components: (i) the predicate $\phi$ describing expected states of the function, and (ii) all the call-site locations where the function is invoked. Thus, we statically extract all precondition check locations through call-graph analysis (\ding{183}) without generating actual specifications, initializing each predicate to the tautological $\true$ as a placeholder. The actual preconditions are generated and refined during Phase 2 (see \cref{sec3-2:interprocedural-spec-syn}). During Phase 1, the \emph{location} of a precondition check is more critical than its predicate content as it determines the structure of the V-Unit queue and thus the order of subsequent verification.

\begin{remark}
The selection of \emph{preconditions} (checked at call sites) as guard assertions, rather than postconditions or loop invariants, is motivated by their intrinsic role in RTE-freeness verification. Preconditions act as natural guards at function entry points, specifying requisite constraints on input parameters (e.g., value ranges, memory accessibility) (see \cref{fig-1:example-analysis-verification}~(c)). In contrast, postconditions and loop invariants capture semantic behaviors of programs or loops (cf.\ \cref{fig-1:example-analysis-verification}~(d) and \cref{fig-3:gen-spec-based-on-feedback}). For RTE-freeness, the goal is to derive (weakest) preconditions that encompass all valid program traces while excluding erroneous ones. %Weakest postconditions or loop invariants (e.g., \texttt{true}) offer no effectiveness for verification.
Consequently, a violated precondition signals a potential RTE, whereas a violated postcondition or invariant is possibly invalid. This focus on preconditions also contributes to the termination guarantee of \ours, as established in \cref{sec:soundness-termination}.
\qedT
\end{remark}

{\ours} deals with acyclic call graphs formally described as follows:

\begin{definition}[Call Graph]\label{def:call-graph}
    A \emph{call graph} is a directed acyclic graph $\cgraph \defeq \langle F, E\rangle$, where
    \begin{itemize}
        \item $F = \{f_1, \dots, f_n\}$ is a finite set of \emph{function nodes},
        \item $E \subseteq F \times F$ is a set of \emph{call edges}, with $\langle f_i, f_j \rangle \in E$ indicating $f_i$ calls $f_j$.
    \end{itemize}%
    The \emph{callees} of a function $f \in F$ are collected into 
    %We define a mapping $\textit{callee}: F\rightarrow \mathcal{P}(F)$ as:
    \begin{align*}
        \textit{callee}\left(f\right) \ddefeq \left\{\,f' \mid \langle f,f' \rangle \in E\,\right\}~.
    \end{align*}%
    Moreover, the \emph{ancestor} relation is defined as
    \begin{multline*}
        \textit{ancestor}\, \ddefeq \left\{\, \left\langle f_\text{anc}, f_\text{des}\right\rangle \in F \times F \,\mid\, \exists f_i,\ldots, f_{j}\colon f_i \in \textit{callee}(f_\text{anc}) \land\right.\\ \left. f_{i+1} \in \textit{callee}(f_i) \land\ldots \land f_j \in \textit{callee}(f_{j-1}) \land f_\text{des} \in \textit{callee}(f_j) \, \right\}.
    \end{multline*}%
    Intuitively, $\langle f_\text{anc}, f_\text{des}\rangle \in \textit{ancestor}$ means that there exists a finite call chain from $f_\text{anc}$ to $f_\text{des}$.
\end{definition}

\begin{figure}[t]
  \centering
  \includegraphics[width=\linewidth]{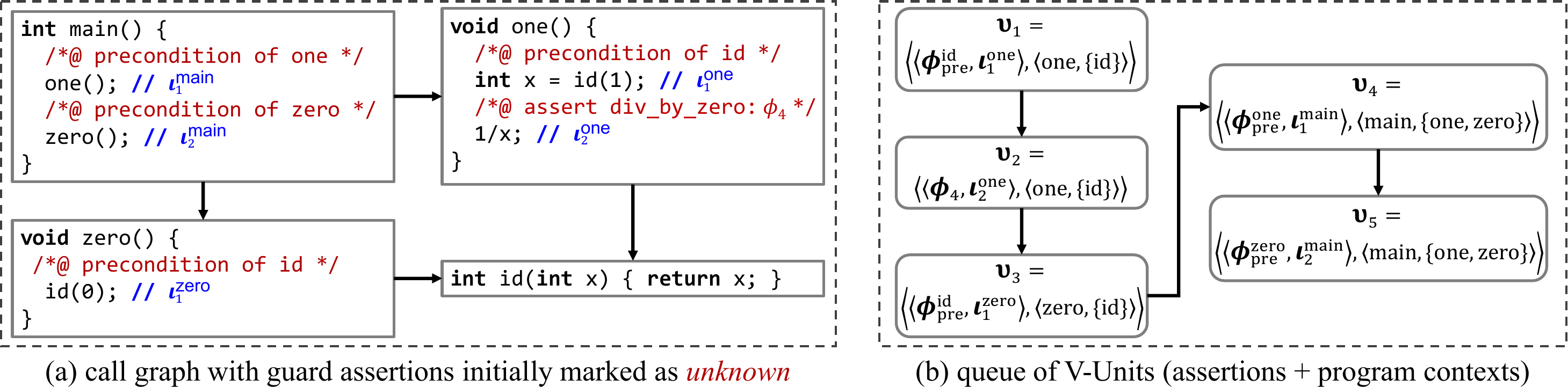}
  \caption{The call graph (a) and V-Unit queue (b) of program \texttt{id.c} in \cref{fig-2:dependencies-spec-property}.}
  \label{fig-5:phase-1-motivating-example}
\end{figure}

\begin{example}
    \cref{fig-5:phase-1-motivating-example}~(a) illustrates the call graph of program \texttt{id.c} in \cref{fig-2:dependencies-spec-property}. The division-by-zero RTE assertion $p_4 = \langle \phi_4, \iota_2^\text{one} \rangle$ as well as all the call-site precondition checks (as placeholders) -- namely $\langle \phi_\text{pre}^\text{id}, \iota_1^\text{one} \rangle$, $\langle \phi_\text{pre}^\text{id}, \iota_1^\text{zero} \rangle$, $\langle \phi_\text{pre}^\text{one},\iota_1^\text{main} \rangle$, and $\langle \phi_\text{pre}^\text{zero}, \iota_2^\text{main} \rangle$ -- are initially marked as $\tagUnknown$. Note that each precondition predicate is initialized to tautological $\true$, which represents the weakest possible constraint, i.e., no semantic restrictions at the initial stage.
    \qedT
\end{example}

% \begin{remark}
% {\ours} admits, in its current implementation, merely \emph{acyclic} call graphs. This is because the underlying deductive verifier \Wp we use lacks comprehensive support for verifying recursive and mutually recursive functions~\cite[Sect.~4.4.2]{frama-c/wp}. We foresee that \ours could be extended to cope with cyclic call graphs by interfacing with verifiers that support recursion while adapting the V-Unit ordering (see \cref{def:v-unit-order}) and the interprocedural specification-synthesis mechanism (see \cref{sec3-2:interprocedural-spec-syn}), though such adaptations may increase the risk of generating over-constrained preconditions for recursive functions (as outlined in \cref{sec:interprocedural-spec}).
% \qedT
% \end{remark}

\begin{remark}
{\ours} admits, in its current implementation, merely \emph{acyclic} call graphs. This is because the underlying deductive verifier \Wp we use lacks comprehensive support for verifying recursive and mutually recursive functions~\cite[Sect.~4.4.2]{frama-c/wp}. Crucially, this acyclicity assumption greatly simplifies our methodological design of the V-Unit ordering (see \cref{def:v-unit-order}) and the interprocedural specification-synthesis mechanism (see \cref{sec3-2:interprocedural-spec-syn}), as it circumvents the complex dependencies in (mutual) recursions and their specifications. We foresee that \ours could be extended to cope with cyclic call graphs by interfacing with verifiers that support recursion while adapting the methodological design, though such adaptations may increase the risk of generating over-constrained preconditions for recursive functions (as outlined in \cref{sec:interprocedural-spec}).
\qedT
\end{remark}

\subsubsection{V-Unit Initialization and Prioritization}\label{sec3-1-2:v-unit}

% \ours leverages the \emph{modularization principle of deductive verification}~\cite[Chap.~9]{deductive-verification}, wherein each function is verified in isolation from the others. Specifically, verifying a function with all its specifications except the preconditions\footnote{As discussed in \cref{sec3-1-1:rte-guard-assertion}, the preconditions of a function are checked at the corresponding call sites, which is actually part of the obligation for verifying its caller functions.} requires only (i) the code implementation of the function, and (ii) the contracts of all callee functions. Therefore, rather than feeding LLMs with the entire program in an end-to-end manner, validating an individual RTE assertion requires a minimal-constrained program context consisting of its host function and all its callee functions, encapsulated within a V-Unit (e.g., \ding{185}). \emph{This design ensures that the context does not expand significantly during the bottom-up verification progression, thereby enabling scalability to large-scale programs.}

\ours leverages the \emph{modularization principle of deductive verification}~\cite[Chap.~9]{deductive-verification}, whereby each function is verified separately. Specifically, to verify that a function adheres to all its specifications except preconditions\footnote{\label{fn:precondition}A function's preconditions are validated at its \emph{call sites}, as part of the verification obligation for its callers (see \cref{sec3-1-1:rte-guard-assertion}).}, one needs \emph{not} the entire program but essentially two elements: (i) the function's own implementation, and (ii) the contracts of all its callees (which are recursively derived from further callees along the call chains in a \emph{bottom-up} manner). Consequently, to validate an individual RTE assertion $\alpha$, it suffices to extract a \emph{minimally constrained program context} consisting of the host function of $\alpha$ and all its callees. \emph{This design prevents the verification context from growing substantially during the bottom-up verification, thereby enabling scalability to large-scale programs.}

{\ours} encapsulates these minimally constrained program contexts into dedicated V-Units:

\begin{definition}[V-Unit]\label{def:v-unit}
    A \emph{V-Unit} is a pair $\langle \alpha,P_\text{context}\rangle$, where $\alpha = \langle \phi,\iota \rangle$ is a guard assertion and $P_\text{context}\defeq\langle f, \textit{callee}(f) \rangle$ is a two-layer program slice containing the host function $f$ of $\alpha$ and its callees.
\end{definition}

{\ours} prioritizes all V-Units into a queue (\ding{184} of \cref{fig-4:framework}) according to a total order $\sqsubseteq$:
\begin{definition}[Order over V-Units]\label{def:v-unit-order}
    Given two V-Units $\upsilon_i = \langle \langle \phi_i, \iota_i \rangle, \langle f_i, \textit{callee}(f_i)  \rangle \rangle$ and $\upsilon_j = \langle \langle \phi_j, \iota_j \rangle, \langle f_j, \textit{callee}(f_j)  \rangle \rangle$, we have $\upsilon_i \sqsubseteq \upsilon_j$ if and only if
    \begin{equation}\label{eq:order}
        \underbrace{\langle f_j, f_i \rangle \in \textit{ancestor}}_{\mathclap{\textnormal{Condition 1: bottom-up verification principle}}}
        ~~\lor~~
        \underbrace{\left(\langle f_j, f_i \rangle \notin \textit{ancestor} ~\land~ \langle f_i, f_j \rangle \notin \textit{ancestor} ~\land~ \textit{loc}(\iota_i) \leq \textit{loc}(\iota_j)\right)}_{\mathclap{\textnormal{Condition 2: total order enforcement}}}~,
    \end{equation}%
    where $\textit{loc}(\iota)$ denotes the location (i.e., line number) of instruction $\iota$.
\end{definition}
\noindent
The design principles underneath the total order $\sqsubseteq$ are embodied in Conditions 1 and 2 of \cref{eq:order}. Condition 1 captures the bottom-up verification progression~\cite{cav24-autospec}: Since verifying a function requires the contracts of its callees, functions higher in the call hierarchy (ancestors) must be verified after their callees. Condition 2 serves two primary purposes. First, for guard assertions $\alpha_i$ and $\alpha_j$ within the same function, it is common that one assertion acts as a hypothesis for the other, e.g., $\mathcal{H} \models \alpha_j$ with $\alpha_i \in \mathcal{H}$. In this case, validating $\alpha_i$ first may generate specifications that facilitate the verification of $\alpha_j$. These hypothesis assertions typically appear at earlier program locations if \texttt{goto} statements and loop structures are disregarded (all properties within loops are mutually dependent and thus receive equal priority). Second, for properties from functions in disjoint call chains (i.e., $\langle f_j, f_i \rangle \notin \textit{ancestor} \land \langle f_i, f_j \rangle \notin \textit{ancestor} \land f_i \neq f_j$), their relative priority is arbitrary. The clause $\textit{loc}(\iota_i) \leq \textit{loc}(\iota_j)$ in Condition 2 ensures that $\sqsubseteq$ is a total order, simplifying its implementation. \cref{fig-5:phase-1-motivating-example}~(b) demonstrates the construction of five V-Units derived from the guard assertions in \cref{fig-5:phase-1-motivating-example}~(a) and their prioritization into a verification queue as per the order $\sqsubseteq$.

% \begin{framed}
% \noindent
% \blue{\textbf{Solution to Challenge 1 in \cref{sec:interprocedural-spec}.} \ours adopts a \emph{Divide} strategy, which decomposes the holistic RTE-freeness verification task into a sequence of manageable V-Units.}
% \end{framed}

% \subsection{Generating Interprocedural Specifications}
\subsection{Phase 2: Fine-Grained Interprocedural Specification Synthesis}
\label{sec3-2:interprocedural-spec-syn}

To address \cref{cha:fine-grained} (fine-grained synthesis), Phase 2 employs a fine-grained strategy, for each V-Unit, to generate interprocedural specifications, as depicted in \cref{fig-4:framework} (lower part). For each V-Unit $\upsilon = \langle \alpha, \langle f, \textit{callee}(f) \rangle \rangle$, \ours utilizes an iterative approach (\ding{186}) comprising two core components: (i) an LLM agent producing specification candidates driven by verifier feedback on guard assertion $\alpha$, and (ii) a verifier validating these candidates both syntactically and semantically, thereby ensuring the soundness of \ours. Each iteration generates interprocedural specifications tailored to the program context $\langle f, \textit{callee}(f) \rangle$ in a refined manner (see details below). Upon successful verification of $\alpha$, \ours advances to the next V-Unit in the queue or terminates if $\upsilon$ is the final unit (\ding{187}). If \ours fails to generate specifications sufficient to validate $\alpha$, the guard assertion is designated as a hypothesis requiring manual review upon termination (\ding{188}). Then, by assuming the validity of $\alpha$, \ours continues processing subsequent V-Units until the queue is exhausted (\ding{189}).

\begin{algorithm}[t]
    \caption{Fine-Grained Interprocedural Specification Synthesis}
    \label{alg:generate-interprocedural-spec}
    \small
    \SetAlCapSkip{1ex}
    \SetKwInput{Input}{Input}\SetKwInput{Output}{Output}\SetKwInput{Parameters}{Parameters}\SetNoFillComment
    %\Input{$\mathcal{A}$: the set of guard assertions, $Q_{\upsilon}$: V-Unit queue, $\textit{prog}$: source program.}
    \Input{~$Q_{\upsilon}$: V-Unit queue, $\textit{prog}$: source program.}
    \Output{~$\mathcal{S}$: synthesized specifications, $\mathcal{V}$: verified properties, $\mathcal{H}$: $\tagUnknown$ hypotheses.}
    \Parameters{~$\text{Verify}$: an extended sound verifier (as per \cref{eq:sound-verifier-extended}), $\textit{iter}$: maximum iterations.}

    $\mathcal{S}\leftarrow \emptyset,~ \mathcal{V}\leftarrow \emptyset,~ \mathcal{H}\leftarrow \emptyset$ \tcp*{initialization} \label{alg:synthesis-initialization}
    \While{$Q_{\upsilon}$ is not empty\label{alg:synthesis-empty-queue}}{
        $\langle \alpha,\langle f,\textit{callee}(f)\rangle\rangle\leftarrow \text{pop}(Q_{\upsilon}),~ \mathcal{S}' \leftarrow \emptyset$ \tcp*{pop the next V-Unit on the queue} \label{alg:synthesis-pop-V-Unit}
        %\BlankLine
        % /* Two-stage interprocedural specification generation. */\\
        \For{$i ~\text{from}~ 1 ~\text{to}~ \textit{iter}$\label{alg:synthesis-refinement-iteration}}{
            % $\langle \text{status},\text{feedback} \rangle \leftarrow \text{Verify}(\textit{prog}, \mathcal{S}\cup\mathcal{S}'\cup\mathcal{A}\setminus\{\alpha\}, \alpha)$\;
            $\langle \text{status},\text{feedback} \rangle \leftarrow \text{Verify}(\textit{prog}, \mathcal{V}\cup\mathcal{H}\cup\mathcal{S}', \alpha)$ \tcp*{apply sound verifier} \label{alg:synthesis-verification}
            \If(\tcp*[f]{refined specification synthesis}){$\text{status}=\tagUnknown$\label{alg:synthesis-refined-synth}}{
                % $\mathcal{S}' \leftarrow \mathcal{S}'\cup\text{generate\_host\_spec}(\alpha, \{f\}, \text{feedback});$ \tcp{Stage 1}
                % $\mathcal{S}' \leftarrow \mathcal{S}'\cup\text{generate\_callees\_spec}(\alpha, f\cup\textit{callee}(f), \text{feedback});$   \tcp{Stage 2}
                $\mathcal{S}' \leftarrow \mathcal{S}'\cup\text{generate\_host\_spec}(\alpha, \{f\}, \mathcal{V}\cup\mathcal{H}, \text{feedback})$\;\label{alg:synthesis-host}
                $\mathcal{S}' \leftarrow \mathcal{S}'\cup\text{generate\_callees\_spec}(\alpha, \{f\}\cup\textit{callee}(f), \mathcal{V}\cup\mathcal{H}, \text{feedback})$\;\label{alg:synthesis-callees}
                $\mathcal{S}' \leftarrow \text{syntax\_and\_semantics\_check}(\mathcal{S}',\textit{prog})$\;\label{alg:synthesis-check}
            }
            \Else{
                $\textbf{break}$\;\label{alg:synthesis-break}
            }
        }
        % \BlankLine
        % \If{$\text{Verify}(\textit{prog}, \mathcal{S}\cup\mathcal{S}'\cup\mathcal{A}\setminus\{\alpha\}, \alpha)=\langle \green{\text{true}}, \_\rangle$}{
        \If(\tcp*[f]{$\alpha$ is validated}){$\text{Verify}(\textit{prog}, \mathcal{V}\cup\mathcal{H}\cup\mathcal{S}', \alpha)=\langle \tagTrue, \_\rangle$}{
            $\mathcal{S} \leftarrow \mathcal{S}\cup\mathcal{S}',~ \mathcal{P} \leftarrow \text{extract\_preconditions}(\mathcal{S})$ \tcp*{collect specifications} \label{alg:synthesis-collect-spec}
            % $\mathcal{A} \leftarrow \mathcal{A}\cup\mathcal{P}, \mathcal{S} \leftarrow \mathcal{S}\cup\mathcal{S}'\setminus\mathcal{P}, Q_\upsilon\leftarrow\text{update}(Q_\upsilon, \mathcal{P})$\;
            $Q_\upsilon\leftarrow\text{update}(Q_\upsilon, \mathcal{P}),~ \mathcal{V} \leftarrow \mathcal{V}\cup\{\alpha\}\cup(\mathcal{S}'\setminus\mathcal{P})$ \tcp*{update $Q_{\upsilon}$ and $\mathcal{V}$ accordingly} \label{alg:synthesis-update}
        }
        \Else{
            $\mathcal{H}\leftarrow \mathcal{H}\cup\{\alpha\}$ \tcp*{augment the hypothesis set with $\alpha$} \label{alg:synthesis-add-hypothesis}
        }
    }
    \Return $\langle \mathcal{S},\mathcal{V},\mathcal{H} \rangle$\;    
\end{algorithm}

% \Cref{alg:generate-interprocedural-spec} details the workflow of Phase 2. The algorithm accepts as input the guard assertions ($\mathcal{A}$) and V-Unit queue ($Q_{\upsilon}$) produced in Phase 1, along with the source program $\textit{prog}$, and yields as output the synthesized specifications $\mathcal{S}$ and the set of \red{unknown} hypotheses $\mathcal{H}$ requiring manual review. Key parameters include an extended sound verifier $\text{Verify}: (\textit{prog},\mathcal{H},q)\mapsto \langle \text{status}, \text{feedback}\rangle$ (cf. \cref{def:verifier}), which returns both verification status and auxiliary feedback, together with hyper-parameters $\textit{iter}_1$ and $\textit{iter}_2$ controlling the maximum refinement iterations for Stage 1 and Stage 2, respectively. The algorithm initializes $\mathcal{S}$ and $\mathcal{H}$ as empty sets (line 1) and processes each V-Unit in $Q_{\upsilon}$ sequentially (lines 2–19). For a V-Unit $\upsilon = \langle a, \langle f,\mathcal{F_\text{callee}} \rangle \rangle$, a temporary specification set $\mathcal{S}'$ is initialized to accumulate newly generated properties (line 3). Stage 1 (lines 5–10) generates preconditions and loop invariants for the host function $f$, while Stage 2 (lines 12–17) produces postconditions for the callee functions $\mathcal{F_\text{callee}}$. If, after both stages, assertion $\alpha$ remains \red{unknown} (line 18), it is added to the hypothesis set $\mathcal{H}$ (line 19). Upon exhausting all V-Units, the algorithm returns the final sets $\mathcal{S}$ and $\mathcal{H}$ (line 20).

\Cref{alg:generate-interprocedural-spec} elaborates the workflow of Phase 2. It takes as input the V-Unit queue $Q_{\upsilon}$ from Phase 1, along with the source program $\textit{prog}$, and yields as output the synthesized specifications $\mathcal{S}$, alongside the verified properties $\mathcal{V}$ and the $\tagUnknown$ hypotheses $\mathcal{H}$ that require manual review. Key parameters to the algorithm include %a sound verifier (\cref{def:verifier}),
an extended sound verifier (see \cref{eq:sound-verifier-extended}) $\text{Verify}\colon (\textit{prog},\mathcal{H},q)\mapsto \langle \text{status}, \text{feedback}\rangle$ -- which returns both verification status and auxiliary feedback --
and the hyper-parameter $\textit{iter}$ that controls the maximum number of refinement iterations. The algorithm initializes $\mathcal{S}$, $\mathcal{V}$ and $\mathcal{H}$ to empty sets (\cref{alg:synthesis-initialization}) and processes each V-Unit in $Q_{\upsilon}$ in sequence (Lines \ref{alg:synthesis-empty-queue}--\ref{alg:synthesis-add-hypothesis}). For each V-Unit $\upsilon = \langle \alpha, \langle f, \textit{callee}(f) \rangle \rangle$, a temporary specification set $\mathcal{S}'$ is initialized to accumulate newly generated properties (\cref{alg:synthesis-pop-V-Unit}). Within the refinement loop (Lines \ref{alg:synthesis-refinement-iteration}--\ref{alg:synthesis-break}), \ours first applies a sound verifier to obtain the status and feedback for the guard assertion $\alpha$ (\cref{alg:synthesis-verification}). If $\alpha$ is $\tagUnknown$ (\cref{alg:synthesis-refined-synth}), \ours leverages the feedback to (i) \emph{generate preconditions and loop invariants for the host function $f$} (\cref{alg:synthesis-host}, detailed in \cref{sec4-2-1:host-func-spec-gen}), and (ii) \emph{generate postconditions for the callee functions $\textit{callee}(f)$} (\cref{alg:synthesis-callees}, detailed in \cref{sec4-2-2:callee-func-spec-gen}). For all these freshly generated specifications, \ours employs a correction and refinement mechanism to enhance their quality and ensure syntactic and semantic validity (\cref{alg:synthesis-check}, detailed in \cref{sec4-2-3:syntax-semantics-check}). \ours exits the loop once $\alpha$ is proven (\cref{alg:synthesis-break}) or the iteration limit $\textit{iter}$ is reached. Finally, if $\alpha$ is certified as \tagTrue, \ours appends the generated specifications $\mathcal{S}'$ to $\mathcal{S}$, and extracts the preconditions $\mathcal{P}$ from the $\mathcal{S}'$ (\cref{alg:synthesis-collect-spec}). %\textcolor{red}{\ours then updates the guard assertions in $Q_\upsilon$ with $\mathcal{P}$, and incorporates the verified $\alpha$ and newly generated specifications except for preconditions $\mathcal{S}'\setminus\mathcal{P}$ into $\mathcal{V}$ (\cref{alg:synthesis-update}).} 
As $\mathcal{P}$ will be validated at $f$'s call sites, \ours then updates the corresponding guard assertions -- which are precondition checks at these call sites -- in $Q_\upsilon$ with $\mathcal{P}$, and incorporates all the newly verified properties -- $\alpha$ and generated specifications $\mathcal{S}'$ except for preconditions $\mathcal{P}$ -- into $\mathcal{V}$ (\cref{alg:synthesis-update}). Otherwise, assertion $\alpha$ is added to the hypothesis set $\mathcal{H}$ (\cref{alg:synthesis-add-hypothesis}).

\subsubsection{Host-Function Specification Generation}\label{sec4-2-1:host-func-spec-gen}

When the verifier fails to validate a guard assertion $\alpha$, \ours leverages LLMs to generate additional specifications by incorporating verifier feedback and the relevant program context, including function $f$ and existing specifications $\mathcal{V}\cup\mathcal{H}$. A simplified version of the prompt we use to instruct the LLM is as follows:

\begin{promptbox}{\small Simplified Prompt for Host-Function Specification Generation}
\small
You are an expert in Frama-C. I will provide a C program specified with ACSL, including an unproven guard assertion, alongside feedback in the form of proof obligations from the {\Wp} verifier.

\textbf{Your tasks:} 
(i) Diagnose the root cause of {\Wp}'s failure to verify the assertion based on the provided feedback; 
(ii) Generate necessary specifications -- such as preconditions, \texttt{assigns} clauses, and loop invariants -- to enable successful verification.

\begin{itemize}
\item \textbf{Program context:} \,\{\(f, \mathcal{V}\cup\mathcal{H}\)\}
\item \textbf{Target guard assertion:} \,\{\(a\)\}
\item \textbf{Proof obligation for the assertion:} \,\{\(\text{feedback}\)\}
\end{itemize}
\end{promptbox}

% \ours adopts a chain-of-thought~\cite{CoT} prompting technique. Specifically, it asks the LLM to first reason about the failure reason of $\alpha$ based on its proof obligation (inspired by Insight 2 in \cref{sec:feedback-driven-spec-syn}), and then generate additional specifications to validate $\alpha$. Note that this is the only chance to generate preconditions for $f$, given only its own program context without the information of call sites where $f$ is invoked, to prevent LLMs misled by part of the call site information to generate over-constrained preconditions, as illustrated in \cref{sec:interprocedural-spec}.

\ours employs a \emph{chain-of-thought} prompting strategy~\cite{CoT}, which instructs the LLM to first reason about the verification failure (\`{a} la \cref{ins:verifier-feedback} on verifier's feedback) before proposing new specifications. Besides, \ours integrates a few specified programs in the prompt as few-shot examples to improve the quality of generated specifications~\cite{few-shot}.
%We observe that such a two-step process enhances logical coherence and corrective relevance. 
Notably, the synthesis is strictly confined to the local context of function $f$, excluding its call-site information to \emph{prevent over-constrained preconditions} -- a critical safeguard as illustrated in \cref{sec:interprocedural-spec}.

\begin{figure}[t]
  \centering
  \includegraphics[width=\linewidth]{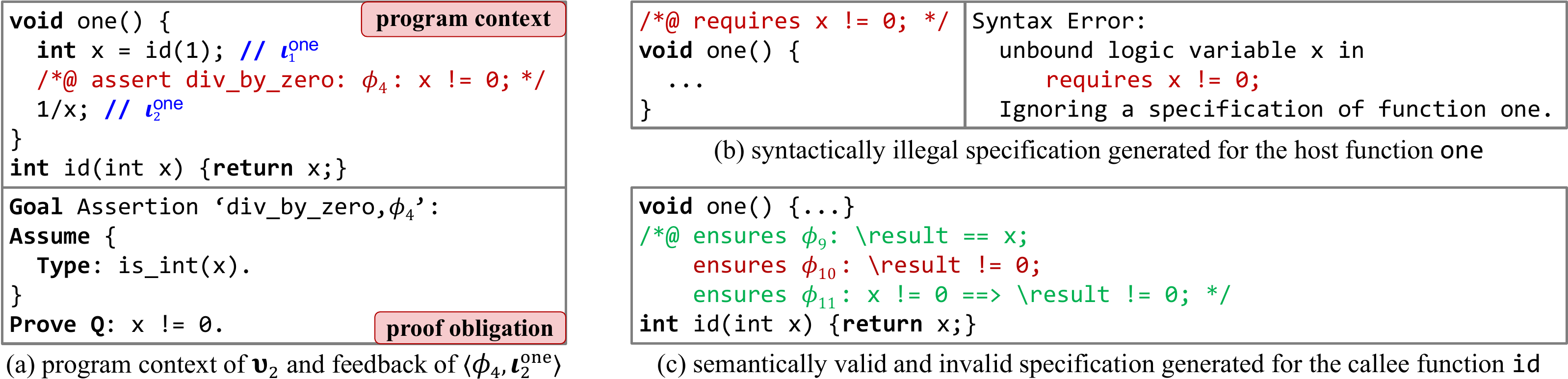}
  \caption{Fine-grained interprocedural specification synthesis for V-Unit $\upsilon_2$ in \cref{fig-5:phase-1-motivating-example}.}
  \label{fig-6:phase-2-motivating-example}
\end{figure}

\begin{example}\label{ex10-host-spec-gen}
    \cref{fig-6:phase-2-motivating-example}~(a) displays the program context of V-Unit $\upsilon_2$ introduced in \cref{fig-5:phase-1-motivating-example}, along with the proof obligation for guard assertion \(\langle \phi_4, \iota_2^\texttt{one} \rangle\). Inspired by the proof goal \textbf{Q}: \(\texttt{x != 0}\), \ours hypothesizes a precondition \red{\texttt{requires x != 0}} for the host function \texttt{one}, as shown in \cref{fig-6:phase-2-motivating-example}~(b). Although this candidate is syntactically illegal initially (as variable \texttt{x} is unbound in the scope of function \texttt{one}), it undergoes an automated correction via the mechanism detailed in \cref{sec4-2-3:syntax-semantics-check}.
    \qedT
\end{example}

\subsubsection{Callee-Functions Specification Generation}\label{sec4-2-2:callee-func-spec-gen}

In this stage, \ours prompts the LLM using a similar approach to that described in \cref{sec4-2-1:host-func-spec-gen}, with two key distinctions: (i) the LLM is provided with the complete program context $\langle f, \textit{callee}(f) \rangle$ (as opposed to only $f$ in the host-function stage), and (ii) the LLM is instructed to generate \emph{postconditions} (and loop invariants) exclusively for the callee functions $\textit{callee}(f)$. This design addresses the scenarios illustrated in \cref{sec:interprocedural-spec}, where validating a guard assertion in function $f$ requires interprocedural specifications -- particularly postconditions -- from its callees. Furthermore, \ours \emph{explicitly prohibits the generation of preconditions} for $\textit{callee}(f)$ to align with the host-function specification mechanism (detailed in \cref{sec4-2-1:host-func-spec-gen}), thereby minimizing the risk of synthesizing over-constrained preconditions.

\begin{example}\label{ex11:callee-spec-gen}
    Consider the program in \cref{fig-6:phase-2-motivating-example}~(c), where the LLM produces three postcondition candidates at this stage: $\langle \phi_9,\iota_\text{post}^\texttt{id} \rangle$, $\langle \phi_{10},\iota_\text{post}^\texttt{id} \rangle$, and $\langle \phi_{11},\iota_\text{post}^\texttt{id} \rangle$. %with predicates $\phi_9 = \green{\texttt{\detokenize{\result == x}}}$, $\phi_{10} = \red{\texttt{\detokenize{\result != 0}}}$, and $\phi_{11} = \green{\texttt{\detokenize{x != 0 ==> \result != 0}}}$. 
    Here, $\langle \phi_9,\iota_\text{post}^\texttt{id} \rangle$ accurately captures the behavior of function \texttt{id}, while $\langle \phi_{10},\iota_\text{post}^\texttt{id} \rangle$ and $\langle \phi_{11},\iota_\text{post}^\texttt{id} \rangle$ are derived from the proof goal \textbf{Q} in \cref{fig-6:phase-2-motivating-example}~(a). Although all three candidates suffice to validate the assertion $\langle \phi_4,\iota_2^\texttt{one} \rangle$, the postcondition $\langle \phi_{10},\iota_\text{post}^\texttt{id} \rangle$ is semantically invalid, which will be screened out as discussed in \cref{sec4-2-3:syntax-semantics-check}.
    \qedT
\end{example}

\subsubsection{Syntactic and Semantic Validity Check of Specifications}\label{sec4-2-3:syntax-semantics-check}
For all the specification candidates generated in \cref{sec4-2-1:host-func-spec-gen,sec4-2-2:callee-func-spec-gen}, \ours employs the verifier to (i) correct syntactically illegal specifications, and (ii) filter out semantically unsatisfiable properties.

When a candidate is syntactically invalid, \ours queries the LLM with verifier-supplied error messages to iteratively correct its syntax. If the candidate remains illegal after several correction attempts, it is discarded. For instance, \cref{fig-6:phase-2-motivating-example}~(b) shows that the precondition \red{\texttt{requires x != 0}} triggers a syntax error due to variable unboundedness. Indeed, any predicate containing variables is invalid for this precondition, as \texttt{one} has no input parameters. Consequently, \ours fails to correct this candidate and hence removes it.

For an $\tagUnknown$ specification that cannot be verified -- and is not a precondition (see \cref{fn:precondition}) -- \ours immediately removes it to avoid inducing false negatives. Consider a pathological example: a loop invariant with predicate \texttt{false} would spuriously validate any property $p$ within the loop, since $\texttt{false} \models p$ trivially holds. To prevent such false negatives, all $\tagUnknown$ non-precondition specifications are discarded. In contrast, an $\tagUnknown$ precondition represents a guard assertion indicating a potential RTE (as discussed in \cref{sec:potential-RTE-Guided-Verif,sec:interprocedural-spec}); it is thus retained in the V-Unit queue to be checked later at the corresponding call sites in caller functions.

\begin{example}
    We illustrate the overall Phase 2 of \ours using the V-Unit queue in \cref{fig-5:phase-1-motivating-example}~(b): The first V-Unit $\upsilon_1$ is verified immediately without additional specifications, as its predicate $\phi_\text{pre}^\texttt{id}$ is initialized to $\true$. \ours then processes $\upsilon_2$, validating it via the postconditions $\langle \phi_9,\iota_\text{post}^\texttt{id} \rangle$ and $\langle \phi_{11},\iota_\text{post}^\texttt{id} \rangle$ generated via steps illustrated in \cref{ex10-host-spec-gen,ex11:callee-spec-gen}. The subsequent V-Units $\upsilon_3$, $\upsilon_4$, and $\upsilon_5$ (pertaining to call sites of \texttt{id}, \texttt{one}, and \texttt{zero}, respectively) are verified immediately since none of these functions require preconditions, and thus no additional specifications are generated.
    \qedT
\end{example}

% \subsection{Design Principles of \ours}\label{sec3-3:rethinking}
% We illustrate the design principles of \ours by discussing several questions.

% \subsubsection{How does \ours solve Challenge 1?}

% \subsubsection{How does \ours solve Challenge 2?}

% \subsubsection{How does \ours guarantee its soundness?}

% \subsubsection{How does \ours guarantee its termination?}

\subsection{Soundness and Termination of \ours}\label{sec:soundness-termination}
This section establishes formal guarantees of {\ours}, including \emph{soundness} and \emph{termination}.

\begin{restatable}[Soundness]{theorem}{restateSoundness}\label{theorem1:soundness}
% \begin{theorem}[Soundness]\label{theorem1:soundness}
    Given a program $\textit{prog}$, suppose {\ours} annotates $\textit{prog}$ with a set of RTE assertions $\mathcal{A}$ and returns synthesized specifications $\mathcal{S}$ together with verified properties $\mathcal{V}$ and $\tagUnknown$ hypotheses $\mathcal{H}$. Then, $\textit{prog}$ is free-of-RTE under hypotheses $\mathcal{H}$ (in the sense of \cref{eq:free-RTE-hypo}).
    % \begin{equation}\label{eq:soundness}
    %     \mathcal{H} \subseteq \mathcal{A} \cup \mathcal{S} \quad \text{and} \quad (\forall p \in \mathcal{A} \cup \mathcal{S} \setminus \mathcal{H}, \quad \mathcal{A} \cup \mathcal{S} \setminus \{p\} \models p).
    % \end{equation}
% \end{theorem}
\end{restatable}

A formal proof of \cref{theorem1:soundness} is provided in \cref{app:proofs}.
Intuitively, if (i) $\mathcal{H}$ is empty or all hypotheses in $\mathcal{H}$ are validated (via manual reviews or auxiliary verification tools), and (ii) the underlying abstract interpretation-based static analyzer is sound (i.e., it misses no potential RTEs), then \ours guarantees that the source program $\textit{prog}$ contains no runtime errors (by \cref{theorem:RTE-freeness}). In that case, all the RTE alarms in $\mathcal{A}$ emitted by the sound analyzer are false positives.

With this soundness assurance, \ours reduces the task of RTE-freeness verification from \emph{reviewing every analyzer-emitted alarm} to \emph{validating each hypothesis generated by \ours}, which, in practice, demands significantly less manual effort; see experimental results in \cref{sec:evaluation}.

In addition to soundness, \ours exhibits termination -- another essential property for practical verification systems. The following termination statement rests on two reasonable assumptions: (i) both abstract interpretation and call-graph analysis terminate successfully, and (ii) every LLM query terminates either naturally or upon reaching a prescribed time limit.

\begin{restatable}[Termination]{theorem}{restateSegmentation}\label{theorem2:termination}
% \begin{theorem}[Termination]\label{theorem2:termination}
    % \ours terminates.
    The \ours procedure always terminates.
% \end{theorem}
\end{restatable}

The following design principles constitute a proof sketch of \cref{theorem2:termination}:

First, the \emph{V-Unit queue is constructed statically} during Phase 1, before the on-the-fly specification synthesis in Phase 2. The queue length is determined solely by the number of RTE assertions generated through abstract interpretation plus the number of call sites identified via call-graph analysis. Although guard assertions in V-Units corresponding to call-site precondition checks may be updated during Phase 2, the number of the V-Units remains invariant.

Second, the verification process exhibits a \emph{one-way progression along the V-Unit queue}: Once a V-Unit is verified or marked as a hypothesis, its validity remains unchanged. As elaborated in \cref{sec4-2-1:host-func-spec-gen,sec4-2-2:callee-func-spec-gen}, for a V-Unit $\upsilon = \langle \alpha, \langle f, \textit{callee}(f) \rangle \rangle$, \ours prohibits the generation of preconditions for callee functions $\textit{callee}(f)$ but only for the host function $f$. Such preconditions for $f$ are then used to update guard assertions in V-Units associated with callers of $f$. By Condition 1 of the ordering $\sqsubseteq$ (see \cref{def:v-unit-order}), these updated V-Units are positioned behind $\upsilon$ in the queue, thereby preserving the validity of all previously verified V-Units.

Third, \emph{the verification of each individual V-Unit is guaranteed to terminate}. This is enforced by the hyper-parameter $\textit{iter}$, which bounds the maximum number of feedback-driven refinement iterations. Moreover, each constituent step -- host-function specification generation, callee-functions specification generation, and syntactic/semantic checking -- is per se designed to terminate.

\subsection{Generality of \ours}\label{sec:generality}

Our \ours framework can be integrated with other sound deductive verifiers supporting formal specification languages, such as OpenJML with JML for Java~\cite{OpenJML} and the Move Prover with MSL for Move~\cite{move}. Although \ours is initially designed for RTE-freeness verification and RTE identification, it can be extended to cope with other vulnerability classes and functional correctness. The key is to substitute RTE assertions with formal annotations flagging the target properties. The subsequent stages, commencing from \ding{183} in \cref{fig-4:framework}, remain fully generic to process these annotations. We prioritize RTE-freeness over full functional correctness because abstract interpretation-based static analysis for generating RTE assertions is mature and sound, while automatically constructing precise functional properties remains challenging \cite{fm21-jcvm-verification}.

\section{Evaluation}\label{sec:evaluation}

The experimental evaluation is designed to address the following research questions (RQ):

\begin{itemize}
    \item[\textbf{RQ1:}] How does \ours compare against state-of-the-art baseline approaches?
    \item[\textbf{RQ2:}] How does \ours perform on large-scale, real-world programs?
    \item[\textbf{RQ3:}] To what extent does each component of \ours contribute to its overall effectiveness?
    \item[\textbf{RQ4:}] How sensitive is \ours to the hyper-parameter $\textit{iter}$ and the underlying LLM?
\end{itemize}

Below, we first report the evaluation setup (\cref{sec:evaluation-setup}), then address \textbf{RQ1}--\textbf{RQ4} (\cref{sec:RQ1,sec:RQ2,sec:RQ3,sec:RQ4}) and conduct case studies (\cref{sec:case-studies}), and finally discuss threats to validity (\cref{subsec:threats}).

\subsection{Evaluation Setup}\label{sec:evaluation-setup}

\paragraph{\bf Implementation} We have implemented {\ours}\footnote{The anonymized artifact for replicating the experimental results is available at \url{https://zenodo.org/records/17296158}.} in about 18,000 lines of Python code, which orchestrates the two-phase divide-and-conquer pipeline, plus 3,000 lines of C++ code for static analysis (e.g., call-graph construction and compilation-context extraction) using Clang. %~\cite{clang}.

We integrate \ours with the deductive verifier \Wp \cite{frama-c/wp} and two abstract interpretation-based static analyzers: \eva \cite{eva-plugin} and \rte \cite{frama-c/rte}. These analyzers exhibit distinct characteristics: \eva is a fine-grained analyzer equipped with advanced abstract domains and data-flow analysis techniques, but it requires the source project to have a unique entry function accompanied by a \emph{generalization mechanism}, e.g., manual assignment of over-approximation value intervals to external variables. In contrast, \rte is a coarse-grained analyzer capable of simultaneously analyzing multiple modules (each with its own entry function), yet often yields more RTE alarms than \eva. 

\paragraph{\bf Configurations}
We repeat all experiments three times and report the averaged results to cope with the inherent uncertainty of LLMs. We use the LLM Claude 4~\cite{claude-4} due to its exceptional performance in program comprehension (API: \texttt{claude-sonnet-4-20250514}; temperature: 0.7; max token: 4,096)\footnote{These settings match those of {\autospec}, with {\autospec} configured to use Claude 4 instead of its original model.} and set the hyper-parameter $\textit{iter}$ (i.e., maximum number of refinement iterations) to 2. For \textbf{RQ4} on sensitivity, we further examine GPT-5~\cite{gpt-5} (API: \texttt{gpt-5-20250807}; temperature: 0.7; max token: 4,096) and vary $\textit{iter}$ from 1 to 5. All the experiments are performed on a MacBook with an Apple M2 chip and 16 GiB RAM, except those for \textbf{RQ1}, which are conducted on a Ubuntu 24.04.2 LTS server with an AMD EPYC 7542 32-core processor and 16 GiB RAM. This is because the baseline approach \autospec (see below) relies on dependencies in Ubuntu.

% \paragraph{\bf Baselines} We compare \ours against \autospec~\cite{cav24-autospec} -- a state-of-the-art LLM-based approach for automated specification generation. Since \autospec assumes that programs are already RTE-free and does not incorporate abstract interpretation, we also evaluate an enhanced version \autospecRte, which integrates the same abstract interpretation-based analyzer (\rte) used in \ours.
% \cref{sec:related-work} discusses the reasons why other previous works can not be considered as fair baselines.

\paragraph{\bf Baselines} We evaluate \ours against two baseline approaches: (i) \autospec~\cite{cav24-autospec}, a state-of-the-art LLM-based method for automated specification generation; and (ii) an enhanced variant, \autospecRte, which integrates the abstract interpretation-based analyzer \rte to generate RTE assertions for the source program before invoking \autospec. Additional rationale for excluding other related works as baselines is provided in \Cref{sec:related-work}.

\paragraph{\bf Benchmarks} We conduct evaluations on one C benchmark suite and four real-world C projects:
\begin{itemize}
    \item Frama-C-Problems~\cite{Frama-c-problems,cav24-autospec} is an open-source benchmark suite used by \autospec, containing 51 small programs with an average of 17.43 lines of code. Each program has an individual function alongside a \texttt{main} entry function. All the programs have been previously verified using {\Wp} with 1$\sim$3 ACSL specifications per program. The rationale for excluding other benchmarks employed by {\autospec} is provided in \cref{app:benchmark}.
    \item Contiki~\cite{contiki} is an open-source operating system for Internet of Things. We evaluate its encryption-decryption module called AES-CCM*, which consists of 544 LoC and 10 functions. Although this module has been manually verified using {\Wp}, the specifications used for its verification are not publicly available.
    \item X509-parser~\cite{x509-parser} is an open-source, RTE-free parser for X.509 format certificates. We evaluate its certificate-policy parsing module, which contains 1,199 LoC and 20 functions. This module has been manually verified with 251 ACSL specifications.
    \item SAMCODE is a spacecraft sun-seeking control system containing 1,280 LoC and 48 functions (not open-sourced due to confidentiality constraints). It has not been previously verified.
    \item Atomthreads~\cite{Atomthreads} is a real-time scheduler for embedded systems. We evaluate its kernel module, which consists of 1,451 LoC and 40 functions. To the best of our knowledge, this module has not been previously verified using {\Wp}.
\end{itemize}

\begin{remark}
To align with industrial verification practices, we adopt the following tactics for large-scale projects in \textbf{RQ2}–\textbf{RQ4}: (i) For SAMCODE, the complete software system with a unique entry function, we manually construct a generalization mechanism and analyze it using \eva; (ii) For Contiki, X509-parser, and Atomthreads, which are system modules requiring substantial manual effort for generalization, we directly apply \rte. For small programs in Frama-C-Problems used for \textbf{RQ1}, \eva alone suffices to establish RTE-freeness without any auxiliary specification or verifier. Therefore, we employ \rte to generate RTE assertions (all of which are false alarms) as verification targets, enabling a comparative evaluation of specification generation and RTE-freeness verification effectiveness between \autospec and \ours.
\qedT
\end{remark}

% \begin{figure*}[t]
%     \begin{minipage}[b]{\textwidth}
%     \begin{table}[H]
%       \centering
%         \caption{Statistics of Benchmarks.}
%         \label{tab:benchmark-statistics}
%         \centering\footnotesize
%         \begin{tabular}{llcccc}
%             \toprule
%             \footnotesize{\textbf{Benchmark Name}} & \footnotesize{\textbf{Type}} & \footnotesize{\textbf{\#Prog}} & \footnotesize{\textbf{(Avg.) LoC}} &
%             \footnotesize{\textbf{\#Func}} & \footnotesize{\textbf{\#Spec}} \\
%             \midrule
%             Frama-C-Problems~\cite{Frama-c-problems,cav24-autospec} & Program suite & 51 & 17.43 & 2 & 1$\sim$3 \\
%             Contiki~\cite{contiki} AES-CCM* & Project module & 1 & 544 & 10 & {--} \\
%             X509-parser~\cite{x509-parser} certificate-policy & Project module & 1 & 1199 & 20 & 251 \\
%             SAMCODE & Entire project & 1 & 1280 & 48 & {--} \\
%             Atomthreads~\cite{Atomthreads} kernel & Project module & 1 & 1451 & 40 & {--} \\
%             \bottomrule
%         \end{tabular}
%     \end{table}
%     \end{minipage}
% \end{figure*}

\subsection{\textbf{RQ1}: Comparison against Baselines}\label{sec:RQ1}

\cref{tab:RQ1} reports the experimental results of \ours against \autospec and \autospecRte. Here, \emph{Num} indicates the number of programs verified as RTE-free (marked as success) in at least one trial across repeated experiments, and \emph{Rate} represents the average success rate per benchmark category. Additionally, we report the average execution time (in minutes) and LLM inference cost (in dollars) for evaluating each approach on the entire Frama-C-Problems benchmark suite. In general, we observe two remarkable disparities between {\ours} and the baselines:

% \begin{figure*}[t]
%     \begin{minipage}[b]{\textwidth}
    \begin{table}[t]
      \centering
        \caption{Experimental result of \ours v.s.\ baselines on Frama-C-Problems.}
        \label{tab:RQ1}
        %\scriptsize
        \setlength{\tabcolsep}{2pt}
        % \begin{tabular}{ccccccccccccccccc}
        \resizebox{\textwidth}{!}{
        \begin{tabular}{lrrrrrrrrrrrrrrrr}
            \toprule
            \multirow{4}{*}{Approach} & 
            \multicolumn{12}{c}{Frama-C-Problems categorized per program features} & 
            \multicolumn{4}{c}{\multirow{3}{*}{\makecell{Overall\\ (51)}} }\\
            \cmidrule(lr){2-13} 
            & 
            \multicolumn{2}{c}{\makecell{General\\ (12)}} & 
            \multicolumn{2}{c}{\makecell{Pointers\\ (8)}} & 
            \multicolumn{2}{c}{\makecell{Loops\\ (8)}} & 
            \multicolumn{2}{c}{\makecell{Arrays\\ (13)}} & 
            \multicolumn{2}{c}{\makecell{Arrays\&Loops\\ (5)}} & 
            \multicolumn{2}{c}{\makecell{Miscellaneous\\ (5)}} & 
            & & & \\
            \cmidrule(lr){2-3}\cmidrule(lr){4-5}\cmidrule(lr){6-7}\cmidrule(lr){8-9}\cmidrule(lr){10-11}\cmidrule(lr){12-13}\cmidrule(lr){14-17}
            & Num & Rate & Num & Rate & Num & Rate & Num & Rate & Num & Rate & Num & Rate & Num & Rate & Time (m) & Cost (\$) \\
            \midrule
            \autospec & 6 & 47.2\% & 6 & 75.0\% & 2 & 25.0\% & 0 & 0.0\% & 1 & 20.0\% & 2 & 40.0\% & 17 & 32.7\% & 287.23 & \textbf{0.448} \\
            \autospecRte & 9 & 69.4\% & \textbf{8} & \textbf{100.0\%} & 1 & 12.5\% & 0 & 0.0\% & 2 & 40.0\% & 2 & 40.0\% & 22 & 41.8\% & 284.40 & 0.455 \\
            \ours & \textbf{12} & \textbf{91.7\%} & \textbf{8} & \textbf{100.0\%} & \textbf{5} & \textbf{37.5\%} & \textbf{11} & \textbf{69.2\%} & \textbf{5} & \textbf{93.3\%} & \textbf{5} & \textbf{100.0\%} & \textbf{46} & \textbf{79.7\%} & \textbf{65.59} & 2.721 \\
            \bottomrule
        \end{tabular}
    }
    \end{table}
%     \end{minipage}
% \end{figure*}

\paragraph{\bf Verification Success Rate} \emph{Overall, \ours successfully verifies 46 out of 51 programs, achieving an average success rate of 79.7\%, which substantially outperforms both \autospec (17/51, 32.7\%) and \autospecRte (22/51, 41.8\%).} Notably, \ours resolves all verification tasks except for a few cases in categories Loops and Arrays, which require complex (nested) loop invariants. These results suggest a promising future improvement of {\ours} by incorporating advanced invariant synthesis techniques, e.g., \cite{ASE24-Wu,ASE24-Pirzada}.

\begin{figure}[t]
  \centering
  \includegraphics[width=\linewidth]{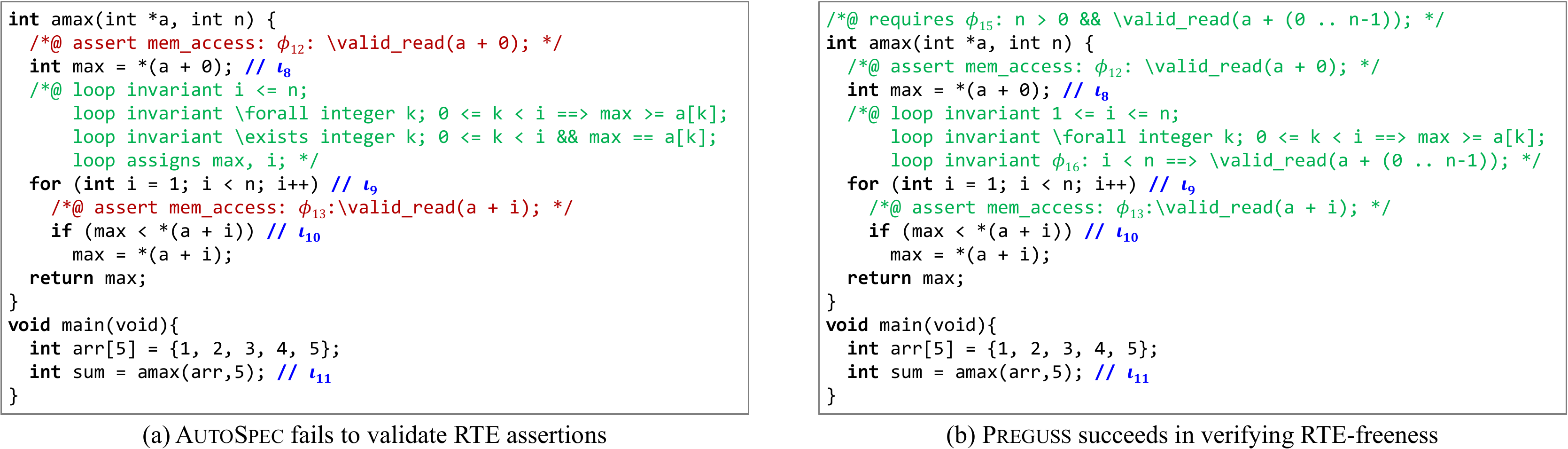}
  \caption{\autospec fails to generate specifications such as $\langle \phi_{15},\iota_\text{pre}^\text{amax} \rangle$ and $\langle \phi_{16},\iota_9 \rangle$ as \ours does, thereby failing to verify the RTE-freeness of \texttt{max.c}.}
  \label{fig:RQ1-study-case}
\end{figure}

It is noteworthy that \autospec underperforms in RTE-freeness verification compared to its functional correctness results (31/51 programs as reported in~\cite[RQ1]{cav24-autospec}) on the same benchmark suite. Besides LLM uncertainty, the primary performance gap lies in the Arrays\footnote{For simplicity, we consolidate the ``immutable\_arrays'', ``mutable\_arrays'', and ``more\_arrays'' categories from~\cite{cav24-autospec} into the single Arrays category.} and Arrays\&Loops categories. Here, \autospec verifies functional correctness for 12/18 cases~\cite{cav24-autospec} but succeeds in RTE-freeness for only 1/18 program (see \cref{tab:RQ1}). This discrepancy is exemplified by program \texttt{max.c} from the Arrays category, as depicted in \cref{fig:RQ1-study-case}. For \texttt{max.c}, which computes the maximum value in an array \texttt{a} of length \texttt{n}, \autospec generates four loop invariants prior to instruction $\iota_9$. Under the assumption that \texttt{max.c} is RTE-free (i.e., $\models \langle \phi_{12},\iota_8\rangle$ and $\models \langle \phi_{13},\iota_{10}\rangle$), as adopted in~\cite[Sect. 4.1]{cav24-autospec}, \autospec produces loop invariants that ensure functional correctness -- the output \texttt{max} indeed represents the array maximum. However, this guarantee presupposes the validity of RTE assertions $\langle \phi_{12},\iota_8 \rangle$ and $\langle \phi_{13},\iota_{10} \rangle$, which are ultimately ensured by the precondition $\langle \phi_{15},\iota_\text{pre}^\text{amax} \rangle$ and loop invariant $\langle \phi_{16},\iota_9 \rangle$ generated by \ours.

The performance gap roots in two fundamental methodological distinctions:
\begin{enumerate}
    \item \textit{Guidance Principle:} \autospec is not RTE assertion-guided; it decomposes the program and synthesizes specifications based solely on program structures -- specifically, an extended call graph where loops and functions are treated as nodes. In contrast, {\ours} uses guard RTE assertions to decompose the holistic verification task into fine-grained V-Units, generating necessary specifications for each. The 29.4\% increase in verified programs (22 by \autospecRte vs.\ 17 by \autospec) underscores the critical role of RTE-assertion guidance.
    \item \textit{Specification-Synthesis Strategy:} The sequence and dependencies of specification generation differ markedly. For instance, in \cref{fig:RQ1-study-case}~(b), {\ours} first generates precondition $\phi_{15}$ to validate guard assertion $\phi_{12}$, then produces loop invariant $\phi_{16}$ at $\iota_9$ for guard assertion $\phi_{13}$ (due to the design principle of the order $\sqsubseteq$ over V-Units; see \cref{def:v-unit-order}). Note that $\phi_{16}$ depends on $\phi_{15}$. {\autospec}, however, prioritizes loop invariants (at $\iota_9$ as the loop is the leaf node in its call graph) without necessary preconditions, leading to invariants such as $\phi_{16}$ being discarded during verification due to their semantic unsatisfiability.
\end{enumerate}

\paragraph{\bf Execution Time and LLM Inference Cost} %Another notable finding is the significant disparity in execution time and LLM inference cost between \ours and the baselines.
Under a 10-minute limit per program, \ours completes all the specification synthesis, whereas \autospec fails to terminate within this limit for 28/51 programs (though it often produces partial output, as seen in \cref{fig:RQ1-study-case}(a) for \texttt{max.c}). This results in \autospec's average execution time being over 4$\times$ longer than that of \ours. Furthermore, \ours consumes approximately 6$\times$ more LLM inference resources than the baselines. This increased cost reflects \ours's strategy of guiding LLMs through a fine-grained synthesis process -- such as employing different strategies upon specification types and incorporating proof obligations -- which yields higher-quality specifications. This aligns with the \emph{test-time scaling paradigm}~\cite{test-time-scaling}, whose viability was recently justified by reasoning models such as OpenAI-o1~\cite{openai-o1}.
% \wzycomment{The test-time scaling means leveraging additional computational resources during the testing (inferring) phase to enhance performance. An interesting observation is: if we calculate the average execution time of \autospec with all timeout cases discarded, then \ours actually spends 4-5x longer execution time on each case, which is consistent with inference cost.} 
Moreover, these high-quality specifications substantially reduce manual verification effort, far outweighing the marginal cost of LLM inference, as demonstrated in \cref{sec:RQ2}.

In summary, \ours demonstrates a substantial advantage over \autospec in verifying RTE-freeness on the Frama-C-Problems benchmark suite. Although \autospecRte improves upon \autospec by incorporating RTE assertion generation, both baselines lack a fine-grained synthesis mechanism that is essential for consistently producing high-quality specifications.

\subsection{\textbf{RQ2}: Evaluation on Large-Scale Real-World Programs}\label{sec:RQ2}

First, we observe that for each of the four large-scale projects, both \autospec and \autospecRte typically fail to terminate within the allotted 15-hour time limit, producing fewer than 15 specifications or aborting unexpectedly. Therefore, we exclude them from the comparison in \textbf{RQ2}.

\begin{table}[t]
  \centering
  \caption{Effectiveness of \ours on large-scale, real-world programs. \#RTE: the number of RTE assertions generated by the static analyzer; Total \#V-Unit: the total number of V-Units constructed; Verified \#V-Unit: the number of verified V-Units (i.e., V-Units with validated guard assertions) across three independent trials; Std.\ \#Spec: the number of specifications required to verify RTE-freeness, which are either sourced from public repositories (X509-parser) or manually crafted by verification experts within a five person-day effort (Contiki and SAMCODE); Generated \#Spec: the number of specifications synthesized by \ours in each experiment; Modified \#Spec: the number of specifications adjusted by experts to achieve full RTE-freeness based on \ours's output, including three types of interventions (from left to right): (i) adding a new property, (ii) correcting an over-constrained precondition, and (iii) removing a property that becomes $\tagUnknown$ after modifying other specifications; Avg.\ SR and Avg.\ HER are explained in \cref{sec:RQ2}.}
    \label{tab:RQ2-effectiveness-of-Preguss}
  \resizebox{\textwidth}{!}{
  \begin{threeparttable} 
    \setlength{\tabcolsep}{3pt}
    % \resizebox{.92\textwidth}{!}{%
    \footnotesize
      \begin{tabular}{lrrrrrrrrrr}
        \toprule
        \footnotesize{\textbf{Benchmark}} &
        \footnotesize{\textbf{\makecell{\#RTE}}} &
        \footnotesize{\textbf{\makecell{Total\\ \#V-Unit}}} &
        \footnotesize{\textbf{\makecell{Verified\\ \#V-Unit}}} &
        \footnotesize{\textbf{\makecell{Avg.\\ SR}}} &
        \footnotesize{\textbf{\makecell{Std.\\ \#Spec}}} &
        \footnotesize{\textbf{\makecell{Generated\\ \#Spec}}} &
        \footnotesize{\textbf{\makecell{Modified\\ \#Spec}}} &
        \footnotesize{\textbf{\makecell{Avg.\\ HER}}} &
        % \footnotesize{\textbf{\makecell{Avg.\\ Time (m)}}} &
        % \footnotesize{\textbf{\makecell{Avg.\\ Cost (\$)}}} \\
        \footnotesize{\textbf{\makecell{Time \\ (m)}}} &
        \footnotesize{\textbf{\makecell{Cost \\ (\$)}}} \\
        \midrule
        Contiki & 163 & 182 & (181, 174, 162) & \textbf{94.7\%} & 111 & (110, \ \ 90, \ \ 96) & (8, 15, 14) & \textbf{88.9\%} & 65.58 & 1.77 \\
        X509-parser & 142 & 174 & (151, 153, 148) & \textbf{86.6\%} & 251 & (108, 116, 142) & (39, 41, 40) & \textbf{84.1\%} & 151.80 & 3.84 \\
        SAMCODE & 44$^\dagger$ & 133 & (121, 116, 119) & \textbf{87.9\%} & 225 & (186, 202, 206) & (34, 59, 38) & \textbf{80.6\%} & 80.48 & 2.36 \\
        % SAMCODE & 44$^\dagger$ & 133 & (121, 116, 119) & \textbf{87.9\%} & 234 & (186, 202, 206) & (50, 72, 79) & \red{71.4\%} & 80.48 & 2.36 \\
        Atomthreads & 239 & 328 & (252, 258, 258) & \textbf{78.1\%} & {--}$^\ddagger$ & (416, 398, 544) & --$^\ddagger$ & {--}$^\ddagger$ & 807.58 & 18.52 \\
        \bottomrule
      \end{tabular}%
    % }%
    \begin{tablenotes}[flushleft]
      \footnotesize
      \item[$\dagger$] We omit floating-point alarms in SAMCODE, which, according to the developers, are not of primary interest.
      \item[$\ddagger$] Std.\ \#Spec, Modified \#Spec, and Avg.\ HER for Atomthreads are unavailable as the manual verification (handcrafting std.\ specifications and refining \ours-generated specifications) was not completed within the 5 person-day limit.
    \end{tablenotes}
  \end{threeparttable}
  }
\end{table}

\Cref{tab:RQ2-effectiveness-of-Preguss} reports the experimental results of \ours applied to large-scale, real-world projects. In addition to metrics described in the table, we compute the \emph{average V-Unit success rate} (Avg.\ SR) and the \emph{average reduction rate of human verification effort} (Avg.\ HER) as follows:
% \begin{equation*}%\label{eq:metrics}
% \small
%     \text{Avg.\ SR} \eeq \frac{\sum{\,\text{Verified \#V-Unit}}}{3\,\times\, \text{Total \#V-Unit}}~,\quad\quad
%     \text{Avg.\ HER} \eeq 1 - \frac{\sum{\,\text{Modified \#Spec}}}{3\,\times\, \text{Std.\ \#Spec}}~.
% \end{equation*}%
% \noindent
\begin{equation*}%\label{eq:metrics}
\small
    \text{Avg.\ SR} \eeq \frac{\sum{\,\text{Verified \#V-Unit}}}{n\,\times\, \text{Total \#V-Unit}}~,\quad
    \text{Avg.\ HER} \eeq 1 - \frac{\sum{\,\text{Modified \#Spec}}}{n\,\times\, \text{Std.\ \#Spec}}
    %,\quad    \revision{n=\text{\#repeated runs (i.e., 3)}}.
\end{equation*}%
where $n = 3$ is the number of repeated experiments accounting for the inherent uncertainty of LLMs. Given the impracticality of fully automating RTE-freeness verification for large-scale projects, Avg.\ SR and Avg.\ HER serve as the primary quantitative measures of \ours's effectiveness. The former reflects the proportion of guard assertions successfully resolved, directing expert attention to the remaining unverified hypotheses. The latter quantifies the reduction in manual effort achieved by \ours compared to verifying the project from scratch.

We observe that \emph{\ours achieves highly automated RTE-freeness verification for the three real-world programs} -- Contiki (544 LoC, 10 functions), X509-parser (1,199 LoC, 20 functions), and SAMCODE (1,280 LoC, 48 functions) -- as reflected by Avg.\ SR and Avg.\ HER both exceeding 80\% with average execution times under 3 hours and LLM inference costs below \$4. These results underscore \ours's effectiveness in synthesizing high-quality specifications and scalability to large projects. In particular, compared to traditional approaches that often require months or even years of expert effort for verifying systems with thousands of LoC \cite{fm21-jcvm-verification,x509-parser}, the computational and financial costs incurred by \ours are marginal. %A typical example of discharging false RTE alarms is given in \Cref{sec:case-studies} (Case~1). 
%(Case~1, drawn from the SAMCODE benchmark).

An interesting observation is that, for the X509-parser benchmark, the total number of generated and modified specifications is substantially lower than the number of standard specifications in all experimental trials. This discrepancy indicates that the ground-truth specifications handcrafted by the X509-parser authors contain many redundant properties -- such as assertions/postconditions describing program states that are not directly relevant to verification, as detailed in \cref{app:RQ2}. This finding further highlights the capability of \ours to produce concise, high-quality specifications.

For the spacecraft control system SAMCODE, besides the 87.9\% automatically verified V-Units, we \emph{identified 6 genuine RTEs} (uninitialized left-value accesses) within the hypothesis set returned by \ours. These errors, which stem from two logical bugs in the system implementation, have been subsequently confirmed by the developers. One such case is detailed in \Cref{sec:case-studies} (Case~2).

An analysis of the manual modifications made to \ours's generated specifications also reveals certain limitations (elaborated in \cref{sec:limitations}). For Contiki, most adjustments involve adding preconditions for C standard library functions such as \texttt{memcpy} and \texttt{memset} from \texttt{string.h}. Since these external functions lack implementation details in the source project, \ours currently relies on handcrafted contracts for them. In X509-parser and SAMCODE, a small number of over-constrained preconditions are generated -- one of which is examined in \Cref{sec:case-studies} (Case~3). After correcting such an over-constrained precondition $p$, properties which are valid under hypotheses including $p$ may become \tagUnknown, and thus should be removed. %, which contributes to the modified \#Spec. 

For the Atomthreads benchmark, \ours achieves an average V-Unit success rate of 78.1\%, albeit with higher resource consumption (average time of 807.58 mins and cost of \$18.52). Resolving the remaining guard assertions is particularly challenging due to the presence of linked-list management modules, which require advanced specifications such as inductive definitions~\cite{frama-c/wp} for full verification (see \cref{sec:limitations}). Constructing these specifications demands prohibitive expert effort -- exceeding our five person-day budget -- % is also difficult for \ours since we do not design special strategies for these advanced specifications.
and falls outside \ours's current design scope. Nevertheless, {\ours} successfully generates high-quality specifications for RTE assertions unrelated to the linked-list module, contributing significantly to the overall success rate.

\paragraph{\bf Specification-Quality Metrics} We evaluate the quality of {\ours}-generated specifications through soundness guarantees and the Avg.\ HER metric, instead of the completeness metrics used in \cite{icse25-specgen,fes24-completeness}. Soundness encompasses (i) \emph{syntactic correctness} of all generated specifications and (ii) \emph{semantic satisfiability} under the hypothesis set $\mathcal{H}$, while Avg.\ HER measures how well {\ours}-generated specifications can substitute for standard oracles. The completeness metrics, which assess how well specifications capture program behaviors for functional correctness verification, are not suitable for {\ours}. This is because {\ours} aims to generate minimal contracts~\cite{minimal-contract} -- the necessary specifications on which guard assertions depend, rather than pursuing the strongest specifications that precisely describe all program behaviors.

\subsection{\textbf{RQ3}: Ablation Study}\label{sec:RQ3}

We conduct an ablation study to assess the individual contributions of key components within the \ours framework, using the large-scale, real-world projects from \textbf{RQ2} (excluding Atomthreads due to its computationally prohibitive verification cost; the same applies to \textbf{RQ4}). Several core modules are deemed essential and thus excluded from ablation:
(i) the \emph{modules for V-Unit construction and prioritization} (\cref{sec3-1:construction-prioritization}), which are fundamental to determining the verification targets establishing the pipeline;
(ii) the \emph{fine-grained specification synthesis strategies} for host and callee functions (\cref{sec4-2-1:host-func-spec-gen,sec4-2-2:callee-func-spec-gen}), as allowing precondition generation for callees would compromise termination guarantees (\cref{sec:soundness-termination}); and
(iii) the \emph{semantic validity check} (\cref{sec4-2-3:syntax-semantics-check}), which prevents false negatives and is critical for soundness.
Consequently, we focus on two adjustable mechanisms: \emph{feedback-driven refinement} (FDR) and \emph{syntax correction} (SC). We evaluate three ablated configurations against the full \ours configuration from \textbf{RQ2}: \ours without FDR (\ours~- FDR), \ours without SC (\ours~- SC), and \ours without both (\ours~- FDR - SC).

\begin{figure}[t]
  \centering
  \includegraphics[width=.9\linewidth]{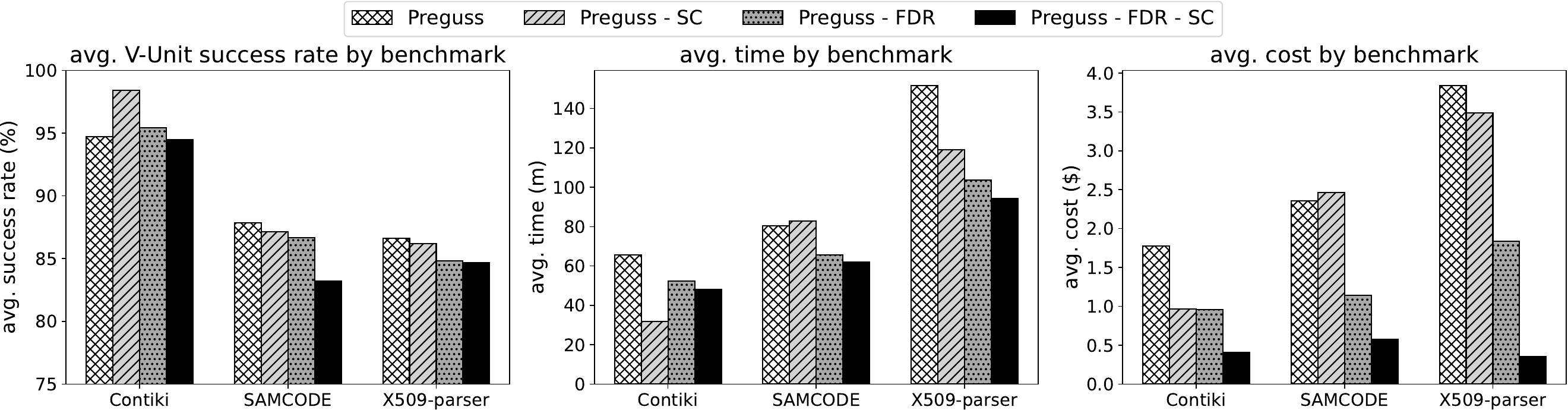}
  \caption{Experimental results of the ablation study.}
  \label{fig:RQ3-ablation-study}
\end{figure}

As illustrated in \cref{fig:RQ3-ablation-study}, \emph{disabling either FDR or SC reduces the average V-Unit success rate, execution time, and LLM inference cost across most benchmarks}, with the combined ablation \ours~- FDR - SC exhibiting the strongest negative effect. An exception occurs in the Contiki benchmark, where the V-Unit success rate unexpectedly increases when SC or FDR is disabled. This anomaly arises because \ours~- FDR - SC achieves a high baseline success rate (94.5\%) on Contiki's relatively small codebase (544 LoC, 10 functions), where LLM uncertainty (temperature = 0.7) dominates over the contributions of FDR and SC. Overall, the high success rates (exceeding 83\%) across all ablated configurations underscore the critical role of \ours's essential modules -- V-Unit management, fine-grained synthesis, and semantic checks -- in achieving effective verification.

% \begin{figure}[t]
%   \centering
%   \includegraphics[width=\linewidth]{figures/RQ4_combined_results_updated-cropped.pdf}
%   \caption{Sensitivity analysis of \ours to the hyper-parameter $\textit{iter}$.}
%   \label{fig:RQ4-hyper-parameter}
% \end{figure}

% \begin{figure}[t]
%   \centering
%   \includegraphics[width=0.86\linewidth]{figures/RQ4-model_comparison_results-cropped.pdf}
%   \caption{Comparative performance of \ours when integrated with Claude 4 v.s.\ GPT-5.}
%   \label{fig:RQ4-llm}
% \end{figure}

\subsection{\textbf{RQ4}: Sensitivity to $\textit{iter}$ and LLMs}\label{sec:RQ4}

% We investigate the effects of the hyper-parameter $\textit{iter}$ (\cref{fig:RQ4-hyper-parameter}) and different LLMs (\cref{fig:RQ4-llm}) on the performance of \ours, evaluated across the Contiki, SAMCODE, and X509-parser benchmarks in terms of V-Unit success rate, execution time, and LLM inference cost. \cref{fig:RQ4-hyper-parameter} indicates \emph{no significant sensitivity of the V-Unit success rate to variations in $\textit{iter}$, whereas execution time and inference cost exhibit a clear increasing trend with higher $\textit{iter}$ values}. The configuration $\textit{iter}=2$ used throughout our other experiments is not finely tuned, yet demonstrates already effective performance. \cref{fig:RQ4-llm} shows that \emph{\ours integrated with Claude 4 achieves a higher V-Unit success rate, reduced execution time, and lower inference cost compared to that with GPT-5}. This performance disparity suggests differences in code comprehension capabilities between the two LLMs.

We observe that \ours exhibits \emph{no significant sensitivity of the V-Unit success rate to variations in $\textit{iter}$, and achieves better performance with Claude 4 compared to GPT-5}; Details are in \cref{app:RQ4}.

\subsection{Case Studies}\label{sec:case-studies}
\begin{figure}[t]
  \centering
  \includegraphics[width=\linewidth]{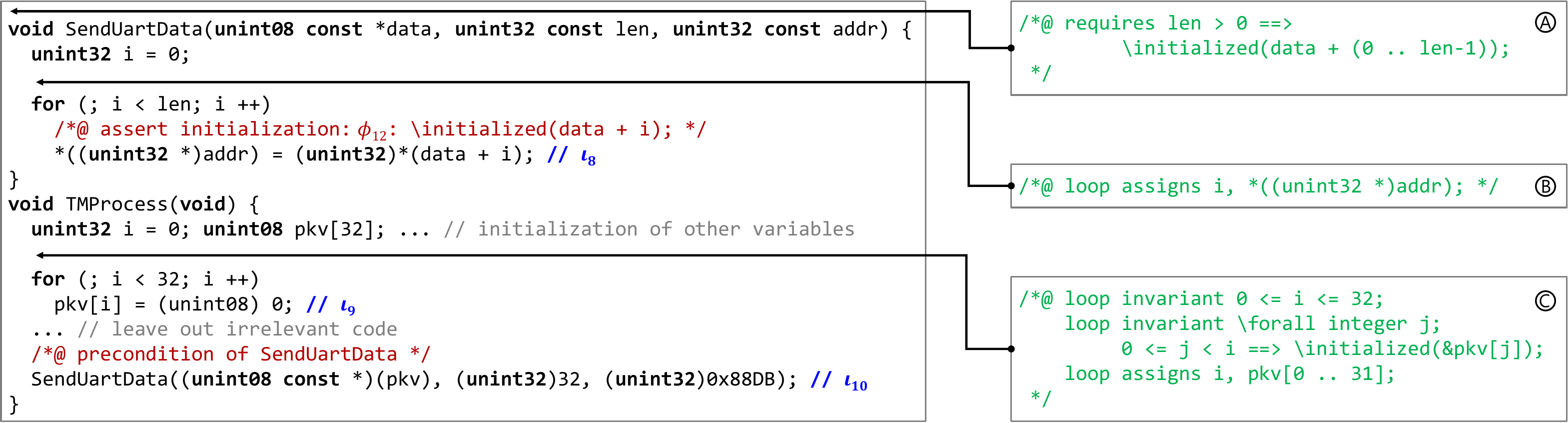}
  \caption{Case 1: discharge false alarm $\langle \phi_{12},\iota_8 \rangle$ through specifications in \Circled{A}--\Circled{C}.}
  \label{fig:study-case-1}
\end{figure}

\begin{figure}[t]
  \centering
  \includegraphics[width=\linewidth]{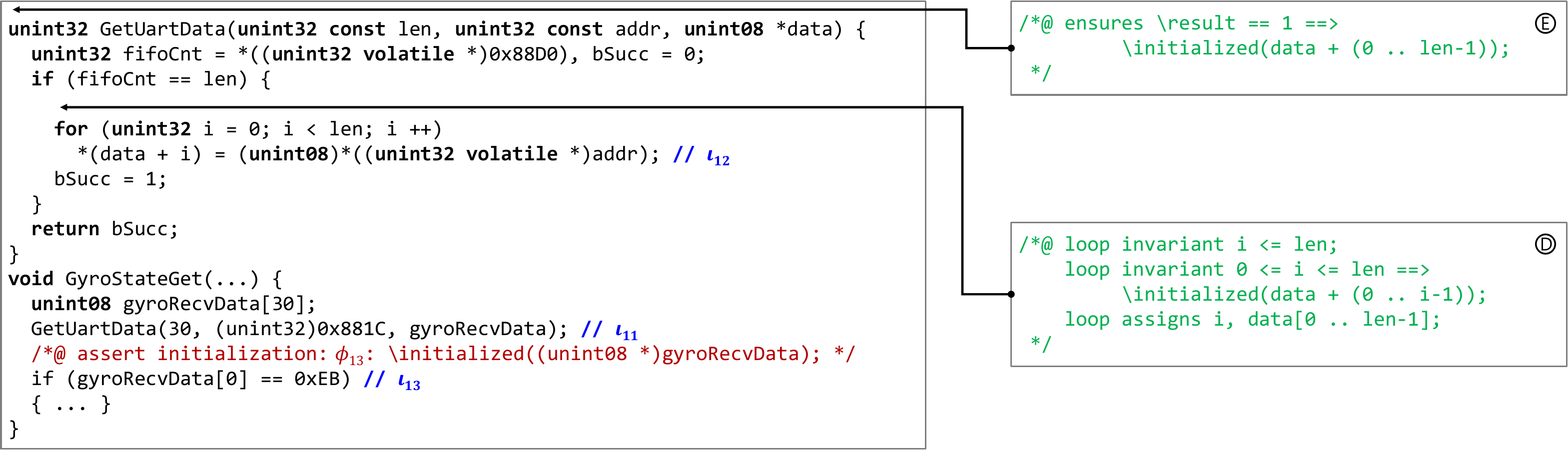}
  \caption{Case 2: specifications in \Circled{D} and \Circled{E} aid in diagnosing the cause of invalid RTE assertion $\langle \phi_{13},\iota_{13} \rangle$.}
  \label{fig:study-case-2}
\end{figure}

\begin{figure}[t]
  \centering
  \includegraphics[width=\linewidth]{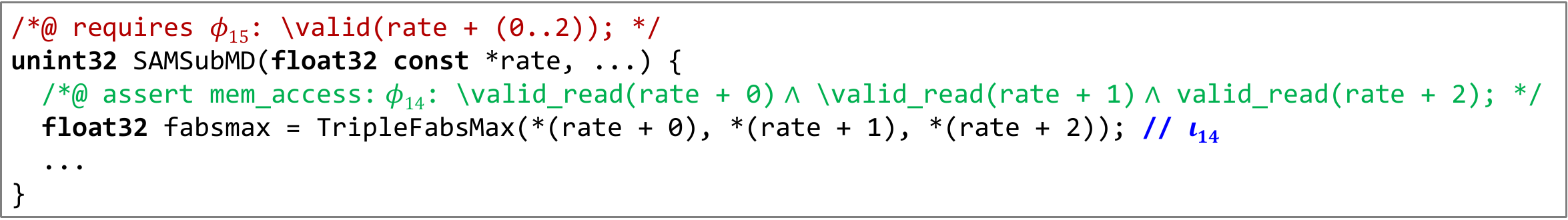}
  \caption{Case 3: the over-constrained precondition $\langle \phi_{15}, \iota_\text{pre}^\text{SAMSubMD} \rangle$ induces a false alarm.}
  \label{fig:study-case-3}
\end{figure}

We now discuss three representative cases found in SAMCODE, the spacecraft sun-seeking control system, to showcase how {\ours} contributes to \emph{discharging false alarms} (\cref{fig:study-case-1}) and \emph{identifying genuine RTEs} (\cref{fig:study-case-2}). We also reveal a situation where \ours may induce false alarms (\cref{fig:study-case-3}). For clarity, we display minimal program contexts and retain only essential specifications.

% \paragraph{\bf Case 1: Discharging False Alarms via Interprocedural Specifications} 
\vspace*{1mm}
\noindent
\textbf{Case 1: Discharging False Alarms via Interprocedural Specifications.}
The program depicted in \cref{fig:study-case-1} consists of two functions: (i) \texttt{SendUartData}, which iteratively transmits each element of array \texttt{data} (length \texttt{len}) to address \texttt{addr} at instruction $\iota_8$; (ii) \texttt{TMProcess}, which initializes each element of the \texttt{uint8} array \texttt{pkv} (length 32) at $\iota_9$ and invokes \texttt{SendUartData} at $\iota_{10}$ to transmit \texttt{pkv} to hardware address \texttt{0x88DB}. The static analyzer generates assertion $p_{12} = \langle \phi_{12}, \iota_8 \rangle$, indicating a potential uninitialized left-value access UB at $\iota_8$. Although $p_{12}$ constitutes a false alarm and could be resolved through exhaustive analyzer configurations (e.g., full loop unrolling), such approaches often incur prohibitive computational costs. Leveraging $p_{12}$ and its verification feedback, \ours synthesizes precondition \Circled{A} and loop invariant \Circled{B} that collectively validate $p_{12}$. Recognizing that the precondition of \texttt{SendUartData} serves as a guard assertion at call site $\iota_{10}$, \ours further generates specifications in \texttt{TMProcess} to discharge this precondition. This yields loop invariants \Circled{C} which guarantee initialization of all \texttt{pkv} elements, and all relevant properties are verified as $\tagTrue$.

% \paragraph{\bf Case 2: Identifying Genuine RTEs via Interprocedural Specifications} 
\vspace*{1mm}
\noindent
\textbf{Case 2: Identifying Genuine RTEs via Interprocedural Specifications.}
The program in \cref{fig:study-case-2} comprises two functions: \texttt{GetUartData} and \texttt{GyroStateGet}. The former first reads value from address \texttt{0x88D0} into variable \texttt{fifoCnt} and then branches as follows: If \texttt{fifoCnt == len} holds, it iteratively receives values from address \texttt{addr}, writes each into the array \texttt{data} (of length \texttt{len}) at instruction $\iota_{12}$, and returns 1; Otherwise, it returns 0. The function \texttt{GyroStateGet} invokes \texttt{GetUartData} at $\iota_{11}$ to initialize array \texttt{gyroRecvData} from hardware address \texttt{0x881C}. The static analyzer flags assertion $p_{13} = \langle \phi_{13}, \iota_{13} \rangle$, indicating a potential uninitialized left-value access UB.

To validate $p_{13}$, \ours generates loop invariants \Circled{D} and a postcondition \Circled{E} for \texttt{GetUartData}, both verified as $\tagTrue$. However, $p_{13}$ remains $\tagUnknown$, suggesting that \texttt{gyroRecvData[0]} may be uninitialized under certain conditions. Manual inspection of postcondition \Circled{E} reveals that \texttt{gyroRecvData} is guaranteed initialized only if \texttt{GetUartData} returns 1. This indicates a latent issue: initialization may fail when the value at address \texttt{0x88D0} does not equal \texttt{len}, causing \texttt{GetUartData} to return 0. To address this, we introduce a conditional check at $\iota_{11}$ to handle the return value of \texttt{GetUartData}, ensuring subsequent instructions execute only when the return value is 1.

% \paragraph{\bf Case 3: Inducing a False Alarm by an Over-Constrained Precondition} 
\vspace*{1mm}
\noindent
\textbf{Case 3: Inducing a False Alarm by an Over-Constrained Precondition.}
The function \texttt{SAMSubMD} in \cref{fig:study-case-3} accepts a set of input parameters, including a constant array \texttt{rate}, and invokes function \texttt{TripleFabsMax} to compute the maximum absolute value of the first three elements of \texttt{rate}. Guided by the analyzer-generated assertion $\langle \phi_{14}, \iota_{14} \rangle$ -- which requires the first three elements of \texttt{rate} to be readable to prevent an invalid memory access UB -- \ours synthesizes a precondition with predicate \red{\texttt{\string\valid(rate + (0..2))}}. This predicate incorrectly asserts that the first three elements of \texttt{rate} must be writable, contradicting the constant nature of \texttt{rate} and thus triggering a false alarm. The correct predicate should be \texttt{\string\valid\_read(rate + (0..2))}. %Although \ours employs refined strategies to mitigate over-constrained preconditions, such issues may still arise occasionally, which demand expert intervention to complete the verification process.
Although \ours employs refined strategies to mitigate over-constrained preconditions, such issues may still arise occasionally, forming a primary component of the hypotheses $\mathcal{H}$ and necessitating expert intervention (e.g., manual weakening of preconditions) to complete the verification.

\subsection{Threats to Validity}\label{subsec:threats}

% \paragraph{Internal Validity}
\noindent
\textit{Internal Validity.}
The first primary threat to internal validity arises from the \emph{inherent uncertainty} in LLM-based inference. To mitigate this, we repeat each experiment three times and report averaged results. The second primary threat is \emph{data leakage}, i.e., are the evaluation code and the corresponding ground-truth ACSL specifications already part of Claude 4's training data? To isolate this potential influence, we build our dataset with three representative real-world projects: (i) Contiki, whose AES-CCM* module is verified but lacks publicly available ACSL specifications; (ii) Atomthreads, which to our knowledge has not been previously verified; and (iii) SAMCODE, which is both unverified and not publicly accessible. The consistent performance of \ours on these benchmarks effectively rules out data-leakage concerns. The third threat concerns the potential misalignment between our evaluation metric -- the V-Unit success rate in \textbf{RQ3} and \textbf{RQ4} -- and the actual reduction in human verification effort (as assessed in \textbf{RQ2}), since the V-Units that remain unvalidated may require more sophisticated specifications than those successfully verified. Nonetheless, given that the primary objectives of \textbf{RQ3} and \textbf{RQ4} are to qualitatively assess the impact of individual components in \ours on its overall performance, V-Unit success rate remains a robust, adequate metric.

\vspace*{1mm}
% \paragraph{External Validity}
\noindent
\textit{External Validity.}
Major threats to external validity concern the soundness of the static analyzer and the completeness of the deductive verifier underpinning \ours. Regarding the former, abstract interpretation-based static analyzers like \rte and \eva typically target specific categories of UBs and RTEs, which may limit the soundness of \ours in guaranteeing RTE-freeness for those categories. For the latter, verifiers such as \Wp rely critically on the underlying SMT solvers (e.g., Z3~\cite{z3} and CVC5~\cite{cvc5}). Limitations in these solvers -- such as inadequate theorem modeling capabilities -- can lead to false alarms where valid properties are incorrectly reported as $\tagUnknown$ hypotheses.
\section{Limitations and Future Directions}
\label{sec:limitations}

We pinpoint scenarios where \ours is inadequate and provide potential solutions thereof.

% Although \ours is initially designed for RTE-freeness verification and genuine RTE identification, it can be extended to cope with other classes of vulnerabilities and functional correctness. The key is to substitute RTE assertions with formal annotations dedicated to the target properties. The subsequent stages, commencing from \ding{183} in \cref{fig-4:framework}, remain fully generic to process these annotations.

\ours currently faces challenges in handling advanced specifications, such as inductive definitions and axioms~\cite{frama-c/wp}, which are especially crucial for verifying programs involving complex data structures like linked lists and trees (recall Atomthreads in \textbf{RQ2}). One potential improvement is to incorporate few-shot examples of such specifications to prompt the LLM.

Despite its refined mechanisms, \ours may occasionally produce over-constrained preconditions due to LLM hallucinations, as illustrated in Case~3 of \Cref{sec:case-studies}. A promising direction is to explore how abstract interpretation-based static analyzers can be used to adjust LLM-generated preconditions by reasoning about over-approximated constraints at call sites. %Additionally, \ours currently lacks a dedicated loop invariant generation mechanism. Integrating advanced loop invariant synthesis techniques could significantly enhance its capability in verifying programs with complex loop structures~\cite{ASE24-Wu,ASE24-Pirzada}.

As project scale increases, the program context of a single V-Unit -- comprising the host function and its callees -- may exceed the LLM's context window limit. Employing program slicing to extract only the callee functions (or the minimal set of statements) that are semantically relevant to the guard assertion could help reduce context size. Moreover, the current design of the program context omits global and external variables, which may result in false alarms related to these variables. Incorporating fine-grained data-flow analysis to retrieve such information could remedy this issue.

Finally, when a project uses external library APIs with unavailable implementations, handcrafted contracts are currently required. A promising extension would be to leverage LLMs to infer plausible contracts from function names and parameter types, thereby reducing the manual effort involved.
\section{Related Work}\label{sec:related-work}

Automated specification synthesis
%, which aims to generate formal descriptions of program behaviors or potential errors, 
remains a critical challenge for scaling program verification~\cite{specification-synthesis}. Traditional techniques are often tailored to specific types of specifications, such as loop invariants~\cite{loop-inv-1,loop-inv-2,loop-inv-3,loop-inv-4}, preconditions~\cite{necessary-preconditions-1,necessary-preconditions-2}, postconditions~\cite{postcondition-1,postcondition-2,postcondition-3}, and assertions~\cite{assertion}. Recently, the emergence of LLMs has led to a new paradigm (cf.\ \cite{zhang2025position}), with several approaches demonstrating significant advantages over traditional methods by leveraging LLMs for specification synthesis~\cite{ICLR24-lemur,ASE24-Wu,ASE24-Pirzada,icse25-specgen,oopsla25-LLM4assertion,cav24-autospec,acl25-cao}. Below, we review closely related LLM-based approaches and clarify their distinctions from \ours.

% \paragraph{LLM-Based Intraprocedural Specification Generation}
\vspace*{1mm}
\noindent
\textit{LLM-Based Intraprocedural Specification Generation.}
Several studies~\cite{ICLR24-lemur,ASE24-Wu,ASE24-Pirzada} aim to generate loop invariants in support of bounded model checkers like ESBMC~\cite{esbmc}, while Laurel~\cite{oopsla25-LLM4assertion} targets assertion synthesis for deductive verifiers such as Dafny. These methods primarily address the synthesis of \emph{intraprocedural properties} within individual functions or lemmas, without considering call relations. In contrast, \ours is designed to generate comprehensive interprocedural specifications -- including contracts and invariants -- for programs with complex call hierarchies. Given this fundamental difference in scope and objective, these intraprocedural approaches are orthogonal to \ours and are thus not selected as baselines.

% \paragraph{LLM-Based Interprocedural Specification Generation}
\vspace*{1mm}
\noindent
\textit{LLM-Based Interprocedural Specification Generation.}
Two notable works, \autospec~\cite{cav24-autospec} and SpecGen~\cite{icse25-specgen}, address interprocedural specification synthesis. Our \ours framework is inspired by \autospec, which employs static analysis to construct a comprehensive call graph (treating both functions and loops as nodes) and iteratively generates specifications for each node in a bottom-up fashion (cf.\ \cref{sec3-1-2:v-unit}). %While \autospec assumes the input programs to be RTE-free and focuses on verifying functional correctness (as illustrated in \cref{fig:RQ1-study-case}), \ours tackles its prerequisite -- the verification of RTE-freeness. Specifically, \ours leverages abstract interpretation to identify potential UBs (flagged by RTE assertions), constructs a prioritized sequence of V-Units, and systematically processes these units in a fine-grained manner. In contrast, 
A thorough discussion of the fundamental methodological distinctions between {\ours} and {\autospec} is presented in \cref{sec:RQ1}.
SpecGen adopts a mutation-based strategy to refine semantically invalid specifications by applying four mutation operators to predicate expressions and selecting the corrected variants. Moreover, SpecGen is implemented on OpenJML~\cite{OpenJML} and targets JML specifications~\cite{jml} for Java programs, whereas \ours is built on \Wp to synthesize ACSL specifications for C programs. %Directly comparing \ours with SpecGen would require substantial engineering effort to adapt to different verification frameworks and specification languages. For these reasons, we select only \autospec as a baseline for comparison.
Given these fundamental differences in verification targets, implementation frameworks, and specification languages, we select only \autospec as a baseline for comparative evaluation.

In a nutshell, \ours demonstrates superior scalability to large-scale programs -- exceeding a thousand lines of code -- due to two key distinctions from existing approaches: (i) it decomposes the monolithic RTE-freeness verification problem into a sequence of %manageable 
V-Units with sliced code contexts,
%, providing LLMs with sliced code contexts rather than the entire programs (as done in prior works), 
thereby mitigating LLM context limitations; and (ii) it generates context-aware interprocedural specifications tailored to the host function and callee functions, rather than producing intraprocedural properties of specific types~\cite{ICLR24-lemur,ASE24-Wu,ASE24-Pirzada,oopsla25-LLM4assertion} or synthesizing all types of specifications indiscriminately~\cite{cav24-autospec,icse25-specgen}.
\section{Conclusion}\label{sec:conclusion}

We have presented {\ours} -- a modular, fine-grained framework for inferring formal specifications synergizing between static analysis and deductive verification. We showcase that {\ours} paves a compelling path towards the automated verification of large-scale programs: It enables nearly fully automated RTE-freeness verification of real-world programs with over a thousand LoC, while facilitating the identification of 6 confirmed RTEs in a practical spacecraft control system.

\section{Data-Avallability Statement}\label{sec:data-availability}

% The artifact (including dataset)~\cite{preguss-artifact} is available at \url{https://doi.org/10.5281/zenodo.18768810}. 
The artifact (including dataset) is available at \url{https://doi.org/10.5281/zenodo.18768810}.

\begin{acks}
    %% acks environment is optional
    %% contents suppressed with 'anonymous'
    %% Commands \grantsponsor{<sponsorID>}{<name>}{<url>} and
    %% \grantnum[<url>]{<sponsorID>}{<number>} should be used to
    %% acknowledge financial support and will be used by metadata
    %% extraction tools.
    %   This material is based upon work supported by the
    %   \grantsponsor{GS100000001}{National Science
        %     Foundation}{http://dx.doi.org/10.13039/100000001} under Grant
    %   No.~\grantnum{GS100000001}{nnnnnnn} and Grant
    %   No.~\grantnum{GS100000001}{mmmmmmm}.  Any opinions, findings, and
    %   conclusions or recommendations expressed in this material are those
    %   of the author and do not necessarily reflect the views of the
    %   National Science Foundation.
    %
    This work has been partially funded by the NSFC under grant No.\ 62572427, by
    the ZJNSF Major Program (No.\ LD24F020013), by the Huawei Cooperation Project (No.\ TC20250422031), by the CCF-Huawei Populus Grove Fund (No.\ CCF-HuaweiSY202503), by the CASC Open Fund (No.\ LHCESET202502), by the NSFC under grant No.\ 62402038 and 62192730, and by the Fundamental Research Funds for the Central Universities of China (No.\ 226-2024-00140). The authors would like to thank Cheng Wen and Yutao Sun for their helpful discussions, Huangying Dong for specifying the SAMCODE spacecraft software, Yazhou Tang for testing the artifact, and the anonymous reviewers for their constructive feedback.
\end{acks}
%
% \wzycommentinline{At most 25 pages for revision and camera-ready version excluding required statements, references, or appendices.}

%% Bibliography file
\bibliography{references}

%% Appendix
\newpage
\appendix
\section*{Appendix}\label{sec:appendix}

\section{Formal Proofs}\label{app:proofs}

Given a source program $\textit{prog}$, \ours first employs a sound abstract interpretation-based static analyzer to generate a set of RTE assertions $\mathcal{A}$. It then returns the synthesized specifications $\mathcal{S}$, alongside verified properties $\mathcal{V}$ and $\tagUnknown$ hypotheses $\mathcal{H}$. 

\subsection{Proof of RTE-Freeness under Valid Hypotheses}\label{sec:appendix-rte-freeness}

% \begin{lemma}[Monotonicity of $\models$]\label{lm:0}
%     \begin{equation}\label{eq:sound-verifier-2}
%         \mathcal{H}\models p \aand \mathcal{H} \subseteq \mathcal{H}' \qimplies \mathcal{H'}\models p~,
%     \end{equation}
% \end{lemma}

% \begin{proof}
% The classification of properties is based on two orthogonal criteria:
% \begin{enumerate}
%     \item \textbf{Generation source}: Any property \(p\) is either generated by the abstract interpretation-based analyzer during Phase~1 (i.e., \(p \in \mathcal{A}\)) or synthesized by the LLM during Phase~2 (i.e., \(p \in \mathcal{S}\)).
%     \item \textbf{Verification status}: Any property \(p\) is either verified as $\tagTrue$ (i.e., \(p \in \mathcal{V}\)) or remains $\tagUnknown$ (i.e., \(p \in \mathcal{H}\)).
% \end{enumerate}
% Since these criteria are mutually exclusive and collectively exhaustive, the equality holds. %\qed
% \end{proof}

\restateRTEFreeness*

% \paragraph{Proof.}
\begin{proof}
To demonstrate that the program satisfies the free-of-RTE condition per \cref{eq:free-RTE}, it suffices to show that every property \(p \in \mathcal{A} \cup \mathcal{S}\) is valid under \(\mathcal{A} \cup \mathcal{S} \setminus \{p\}\) (i.e., \(\mathcal{A} \cup \mathcal{S} \setminus \{p\} \models p\)). We proceed by case analysis on the membership of \(p\) in \(\mathcal{H}\).

\begin{enumerate}
    \item \textbf{Case:} \(p \in \mathcal{A} \cup \mathcal{S} \setminus \mathcal{H}\). By the definition of free-of-RTE under hypotheses \cref{eq:free-RTE-hypo}, we directly obtain:
        \[
        \mathcal{A} \cup \mathcal{S} \setminus \{p\} \models p.
        \]
    \item \textbf{Case:} \(p \in \mathcal{H}\). Since every hypothesis \(h \in \mathcal{H}\) is valid by assumption, the specific hypothesis \(p\) is valid (i.e., \(\models p\)). Consequently, by the definition of $\models$ (\cref{def:validity}), \(p\) is valid by any set of premises, and in particular:
        \[
        \mathcal{A} \cup \mathcal{S} \setminus \{p\} \models p.
        \]
\end{enumerate}

Since the entailment holds in both cases, we conclude that \(\forall p \in \mathcal{A} \cup \mathcal{S}, \mathcal{A} \cup \mathcal{S} \setminus \{p\} \models p\), thereby satisfying the free-of-RTE criterion in \cref{eq:free-RTE}. %\qed
\end{proof}

\subsection{Proof of Soundness}\label{sec:appendix-soundness}

\begin{lemma}\label{lm:1}
    The set of all properties is partitioned such that \(\mathcal{A} \cup \mathcal{S} = \mathcal{V} \cup \mathcal{H}\), where \(\mathcal{A} \cap \mathcal{S} = \emptyset\) and \(\mathcal{V} \cap \mathcal{H} = \emptyset\).
\end{lemma}

% \paragraph{Proof.}
\begin{proof}
The classification of properties is based on two orthogonal criteria:
\begin{enumerate}
    \item \textbf{Generation source}: Any property \(p\) is either generated by the abstract interpretation-based analyzer during Phase~1 (i.e., \(p \in \mathcal{A}\)) or synthesized by the LLM during Phase~2 (i.e., \(p \in \mathcal{S}\)).
    \item \textbf{Verification status}: Any property \(p\) is either verified as $\tagTrue$ (i.e., \(p \in \mathcal{V}\)) or remains $\tagUnknown$ (i.e., \(p \in \mathcal{H}\)).
\end{enumerate}
Since these criteria are mutually exclusive and collectively exhaustive, the equality holds. %\qed
\end{proof}

% \begin{lemma}\label{lm:2}
%     For every RTE assertion \(a \in \mathcal{A}\), either \(a \in \mathcal{H}\) or \(a \in \mathcal{V}\).
% \end{lemma}

% \paragraph{Proof.}
% This follows directly from \cref{lm:1}, as \(\mathcal{A} \subseteq \mathcal{A} \cup \mathcal{S} = \mathcal{V} \cup \mathcal{H}\). Algorithmically, this is enforced by lines 12–16 of \cref{alg:generate-interprocedural-spec}, where each \(a\) is assigned to \(\mathcal{V}\) or \(\mathcal{H}\) based on verification outcomes. \qed

\begin{lemma}\label{lm:2}
    For any RTE assertion \(a \in \mathcal{A} \cap \mathcal{V}\), the entailment \(\mathcal{V} \cup \mathcal{H} \setminus \{a\} \models a\) holds. This signifies that every RTE assertion classified into \(\mathcal{V}\) is valid under the hypotheses \(\mathcal{V} \cup \mathcal{H} \setminus \{a\}\).
\end{lemma}

% \paragraph{Proof.}
\begin{proof}
Consider an arbitrary \(a \in \mathcal{A} \cap \mathcal{V}\) and its associated V-Unit \(\upsilon_i\). Let \(\mathcal{V}_{i-1}\) and \(\mathcal{H}_{i-1}\) denote the verified properties and $\tagUnknown$ hypotheses prior to processing \(\upsilon_i\), and let \(\mathcal{S}'_i\) be the specifications generated during \(\upsilon_i\)'s verification (lines 4–11 of \cref{alg:generate-interprocedural-spec}). The verification step (line 12) ensures:
\[
\textit{Verify}(\textit{prog}, \mathcal{V}_{i-1} \cup \mathcal{H}_{i-1} \cup \mathcal{S}'_i, a) = \tagTrue,
\]
with \(a \notin \mathcal{V}_{i-1} \cup \mathcal{H}_{i-1} \cup \mathcal{S}'_i\). By the soundness condition of the verifier (cf. \cref{eq:sound-verifier-1}), this implies:
\begin{equation}\label{eq:8}
    \mathcal{V}_{i-1} \cup \mathcal{H}_{i-1} \cup \mathcal{S}'_i \models a.
\end{equation}

Observe that \(\mathcal{S}'_i \subseteq \mathcal{S}\), \(\mathcal{V}_{i-1} \subseteq \mathcal{V}\), and \(\mathcal{H}_{i-1} \subseteq \mathcal{H}\), while \(a \notin \mathcal{S}'_i \cup \mathcal{V}_{i-1} \cup \mathcal{H}_{i-1}\) (since \(a\) is neither generated during \(\upsilon_i\)'s verification nor previously classified). Thus applying \cref{lm:1}:
\begin{equation}\label{eq:9}
    \mathcal{V}_{i-1} \cup \mathcal{H}_{i-1} \cup \mathcal{S}'_i \subseteq \mathcal{V}\cup\mathcal{H}\cup\mathcal{S}\setminus\{a\} =  \mathcal{V} \cup \mathcal{H} \setminus \{a\}.
\end{equation}

Applying \cref{eq:sound-verifier-2} (monotonicity of \textit{Verify}) to \eqref{eq:8} and \eqref{eq:9} yields \(\textit{Verify}(\textit{prog}, \mathcal{V} \cup \mathcal{H} \setminus \{a\}, a) = \tagTrue,\), and thus \(\mathcal{V} \cup \mathcal{H} \setminus \{a\} \models a\) holds. %\qed
    
\end{proof}

\begin{lemma}\label{lm:3}
    For any property \(p \in \mathcal{V}\), the entailment \(\mathcal{V} \cup \mathcal{H} \setminus \{p\} \models p\) holds. This establishes that every verified property remains valid under the set of all other verified properties and hypotheses.
\end{lemma}

% \paragraph{Proof.}
\begin{proof}
Similar to the proof of \cref{lm:2}, consider an arbitrary property \(q \in \mathcal{V}\setminus\mathcal{A}\). By the verification process detailed in \cref{sec4-2-3:syntax-semantics-check}, \(q\) is confirmed to be syntactically and semantically valid through a series of refinement steps. Specifically, during the verification of the corresponding V-Unit, we have \(\textit{Verify}(\textit{prog}, \mathcal{V}_{i-1} \cup \mathcal{H}_{i-1}, q) = \tagTrue\). Since \(\mathcal{V}_{i-1} \cup \mathcal{H}_{i-1} \subseteq \mathcal{V} \cup \mathcal{H} \setminus \{q\}\) (since $q$ is not generated in verification of previous V-Units), the monotonicity of \textit{Verify} (cf. \cref{eq:sound-verifier-2}) guarantees that \(\mathcal{V} \cup \mathcal{H} \setminus \{q\} \models q\) holds.

Then combining \cref{lm:2}, \(\mathcal{V} \cup \mathcal{H} \setminus \{p\} \models p\) holds for any property \(p \in \mathcal{V}\). %\qed
\end{proof}

% \begin{theorem}[Soundness]
%     The source program $\textit{prog}$ is guaranteed to be free of RTEs under hypotheses $\mathcal{H}$, as formally defined by the condition:
%     \begin{equation*}
%         \mathcal{H} \subseteq \mathcal{A} \cup \mathcal{S} \quad \text{and} \quad (\forall p \in \mathcal{A} \cup \mathcal{S} \setminus \mathcal{H}, \quad \mathcal{A} \cup \mathcal{S} \setminus \{p\} \models p).
%     \end{equation*}
% \end{theorem}

\restateSoundness*

% \paragraph{Proof.}
\begin{proof}
% The proof demonstrates that \ours satisfies the two conditions of \cref{eq:soundness}:

% \begin{enumerate}
%     \item \(\mathcal{H} \subseteq \mathcal{A} \cup \mathcal{S}\): This follows directly from \cref{lm:1}.%, which partitions the set of all properties such that \(\mathcal{V} \cup \mathcal{H} = \mathcal{A} \cup \mathcal{S}\). Since \(\mathcal{V} \cap \mathcal{H} = \emptyset\), it is immediate that \(\mathcal{H} \subseteq \mathcal{A} \cup \mathcal{S}\).
    
%     \item Consider an arbitrary \(p\in\mathcal{A}\cup\mathcal{S}\setminus\mathcal{H}\). By \cref{lm:1}, \(p \in \mathcal{V}\) because \(\mathcal{A} \cup \mathcal{S} \setminus \mathcal{H} \subseteq \mathcal{V}\). Applying \cref{lm:3} yields \(\mathcal{V} \cup \mathcal{H} \setminus \{p\} \models p\). Since \cref{lm:1} implies \(\mathcal{V} \cup \mathcal{H} = \mathcal{A} \cup \mathcal{S}\), we have \(\mathcal{V} \cup \mathcal{H} \setminus \{p\} = \mathcal{A} \cup \mathcal{S} \setminus \{p\}\), concluding that \(\mathcal{A} \cup \mathcal{S} \setminus \{p\} \models p\).
% \end{enumerate}

% Thus, both conditions of \cref{eq:free-RTE-hypo} are satisfied. %\qed

Consider an arbitrary \(p\in\mathcal{A}\cup\mathcal{S}\setminus\mathcal{H}\). By \cref{lm:1}, \(p \in \mathcal{V}\) because \(\mathcal{A} \cup \mathcal{S} \setminus \mathcal{H} \subseteq \mathcal{V}\). Applying \cref{lm:3} yields \(\mathcal{V} \cup \mathcal{H} \setminus \{p\} \models p\). Since \cref{lm:1} implies \(\mathcal{V} \cup \mathcal{H} = \mathcal{A} \cup \mathcal{S}\), we have \(\mathcal{V} \cup \mathcal{H} \setminus \{p\} = \mathcal{A} \cup \mathcal{S} \setminus \{p\}\), concluding that \(\mathcal{A} \cup \mathcal{S} \setminus \{p\} \models p\).
\end{proof}

\section{Rationale for not Employing the Full Dataset of {\autospec}}\label{app:benchmark}

The unemployed datasets fall into two categories:
\begin{enumerate}
    \item \emph{Six functions extracted from X509-parser}: Although {\autospec} evaluates these as six separate small-scale verification tasks (each with a handcrafted \texttt{main} function), we consider the X509-parser submodule used in {\ours} to be more representative. This submodule is larger (over 1,000 LoC with 20 functions) and exhibits a more realistic call hierarchy (with a single entry point). Besides, 2 of the 6 functions evaluated by {\autospec} are included in the X509-parser submodule used by {\ours}.
    \item \emph{SyGuS, OOPSLA-13, and SV-COMP benchmarks}: These suites consist primarily of single-function programs that either (i) require no preconditions (due to absent input parameters) or (ii) need only very simple preconditions. Moreover, such single-function programs lack call sites for precondition checks and do not require postconditions. These benchmarks are designed for evaluating intraprocedural specification generation (e.g., loop invariants), whereas {\ours} targets interprocedural specification synthesis, which is essential for verifying medium- to large-scale programs. A detailed rationale is provided in \cref{sec:related-work}.
\end{enumerate}

\section{Details of Redundant Specifications in the Standard Set for \textbf{RQ2}}\label{app:RQ2}

By comparing with {\ours}-generated specifications, we identified two primary types of redundancies in the standard specification set: (i) 40 assertions (e.g., \texttt{assert len >= rbytes}) manually introduced by developers as checkpoints to record intermediate program states, and (ii) a large portion of the 80 postconditions -- compared to the 23, 17, and 35 postconditions generated by {\ours} in three repeated experiments. Although these assertions and postconditions correctly describe certain program behaviors, they do not directly contribute to validating guard assertions.

\begin{figure}[t]
  \centering
  \includegraphics[width=\linewidth]{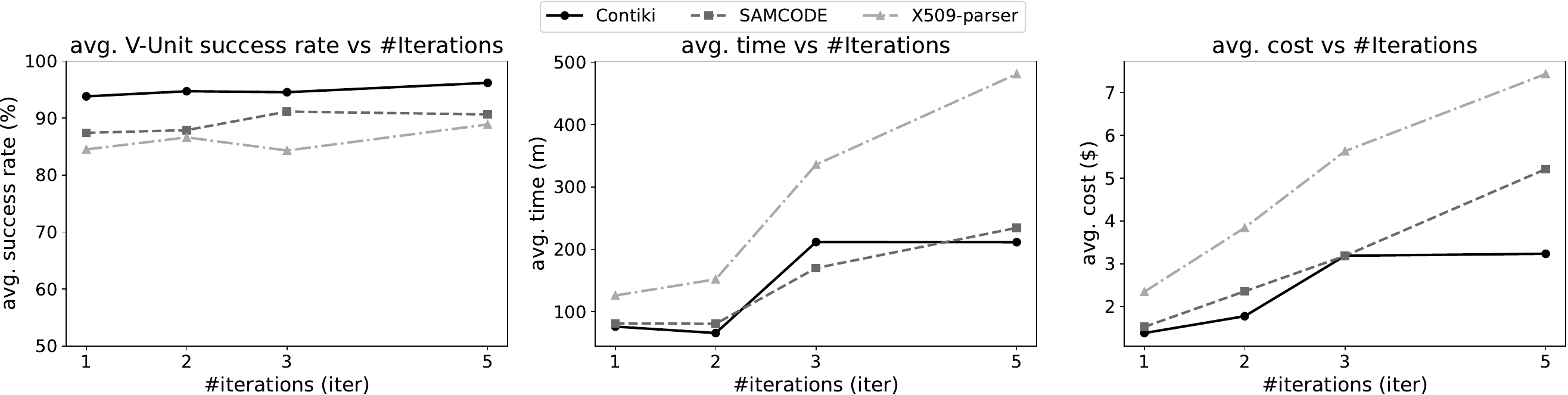}
  \caption{Sensitivity analysis of \ours to the hyper-parameter $\textit{iter}$.}
  \label{fig:RQ4-hyper-parameter}
\end{figure}

\begin{figure}[t]
  \centering
  \includegraphics[width=\linewidth]{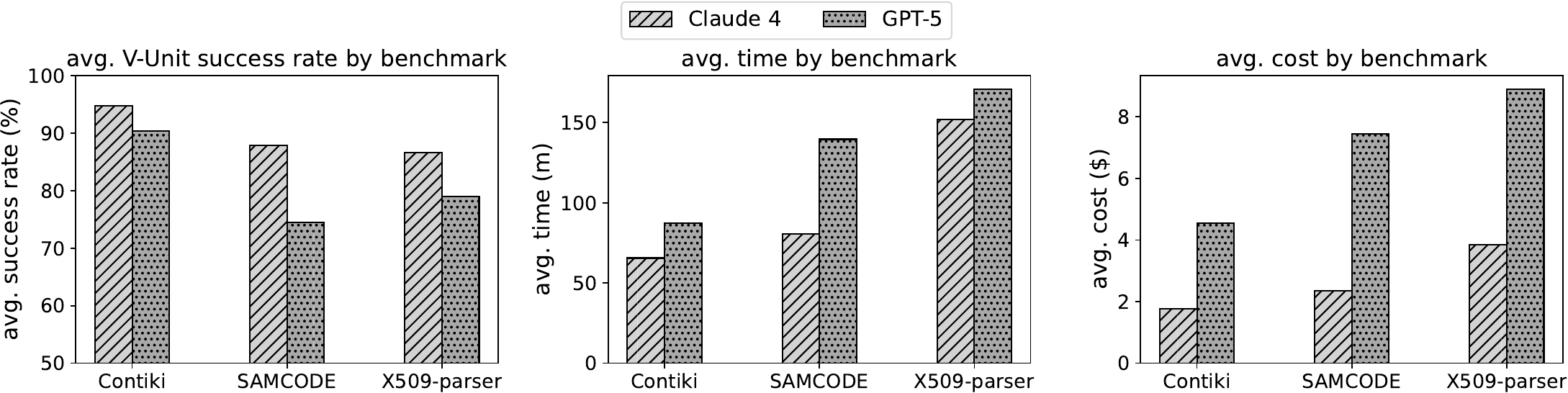}
  \caption{Comparative performance of \ours when integrated with Claude 4 v.s.\ GPT-5.}
  \label{fig:RQ4-llm}
\end{figure}

\section{Details of \textbf{RQ4}}\label{app:RQ4}

We investigate the effects of the hyper-parameter $\textit{iter}$ (\cref{fig:RQ4-hyper-parameter}) and different LLMs (\cref{fig:RQ4-llm}) on the performance of \ours, evaluated across the Contiki, SAMCODE, and X509-parser benchmarks in terms of V-Unit success rate, execution time, and LLM inference cost. \cref{fig:RQ4-hyper-parameter} indicates \emph{no significant sensitivity of the V-Unit success rate to variations in $\textit{iter}$, whereas execution time and inference cost exhibit a clear increasing trend with higher $\textit{iter}$ values}. The configuration $\textit{iter}=2$ used throughout our other experiments is not finely tuned, yet demonstrates already effective performance. \cref{fig:RQ4-llm} shows that \emph{\ours integrated with Claude 4 achieves a higher V-Unit success rate, reduced execution time, and lower inference cost compared to that with GPT-5}. This performance disparity suggests differences in code comprehension capabilities between the two LLMs.

% \section{Complete Prompt}
% \wzycommentinline{Give the complete version of LLM prompt.}

\end{document}